%% file: raddiag_submit.tex
\documentclass[natbib,twocolumn]{svjour3}                     
\smartqed  
 \usepackage{mathptm}      

\usepackage{amsmath,amsfonts,amstext,mathrsfs,float,tabu,array}
\usepackage{mathtools}
\usepackage{times}
\usepackage[official]{eurosym}

\usepackage{bbm}
\usepackage{twoopt}
\usepackage{rotating}	
\newcommand\ion[2]{#1$\,${\scshape{#2}}}
\usepackage{aas_macros}
\def\mathbi#1{\textbf{\em #1}}

\newcommand {\kv} {$\mathbf{K}$}

\newcommand\Tstrut{\rule{0pt}{4ex}}       
\newcommand\Bstrut{\rule[-3ex]{0pt}{0pt}} 
\newcommand{\TBstrut}{\Tstrut\Bstrut} 
 \usepackage{mathptm}      
 \usepackage{times}
\makeatletter
\newcommandtwoopt{\citeads}[3][][]{\href{http://adsabs.harvard.edu/abs/#3}%
{\def\hyper@linkstart##1##2{}%
\let\hyper@linkend\@empty\citealp[#1][#2]{#3}}}
\newcommandtwoopt{\citepads}[3][][]{\href{http://adsabs.harvard.edu/abs/#3}%
{\def\hyper@linkstart##1##2{}%
\let\hyper@linkend\@empty\citep[#1][#2]{#3}}}

\newcommandtwoopt{\citetads}[3][][]{\href{http://adsabs.harvard.edu/abs/#3}%
{\def\hyper@linkstart##1##2{}%
\let\hyper@linkend\@empty\citet[#1][#2]{#3}}}
\newcommandtwoopt{\citeyearads}[3][][]%
{\href{http://adsabs.harvard.edu/abs/#3}
{\def\hyper@linkstart##1##2{}%
\let\hyper@linkend\@empty\citeyear[#1][#2]{#3}}}
\makeatother

\begin{document}

\title{Radiative diagnostics in the solar photosphere and chromosphere}

\titlerunning{Radiative diagnostics in the solar photosphere and chromosphere}        

\author{J.~de~la~Cruz~Rodr\'iguez \and M.~van~Noort}

\institute{J. de la Cruz Rodr\'iguez \at
  Institute for Solar Phyics, Stockholm University \\Albanova
  University Center, SE-10691, Stockholm, Sweden\\
              \email{jaime@astro.su.se}           
           \and
           M. van Noort \at
           Max Planck Institute for Solar System Research \\Justus-von-Liebig-Weg 3, 37077 G\"ottingen, Germany\\
           \email{vannoort@mps.mpg.de}
}

\date{Received: date / Accepted: date}

\maketitle

\begin{abstract}
Magnetic fields on the surface of the Sun and stars in general imprint or modify the polarization state of the electromagnetic radiation that is leaving from the star. The inference of solar/stellar magnetic fields is performed by detecting, studying and modeling polarized light from the target star. In this review we present an overview of techniques that are used to study the atmosphere of the Sun, and particularly those that allow to infer magnetic fields. We have combined a small selection of theory on polarized radiative transfer, inversion techniques and we discuss a number of results from chromospheric inversions.

\end{abstract}

\section{Introduction}

\input{intro.tex}

\section{Radiative transfer and Response functions}\label{sec:rad}
\input{response.tex}

\section{Inversion methods for solar observations}\label{sec:mod}
\input{model.tex}

\input{obs.tex}

\section{Discussion and future developments}\label{sec:fut}
\input{future.tex}

\begin{acknowledgements}
J. de la Cruz Rodr\'iguez acknowledges financial support from Vetenskapsr\aa det (VR, the Swedish Research Council) and the Swedish National Space Board (SNSB).
\end{acknowledgements}


\end{document}

%% file: intro.tex
Stellar spectra encode information of the composition and of the physical conditions present in the hosting star where radiation originates. Compared to other research fields in physics, astronomy is mostly a \emph{remote sensing} science. Solar and stellar physics have developed through the study of radiation and polarization, attempting to infer the underlying physical state of the atmospheric plasma. While most stellar observations are spatially unresolved\footnote{See latest advances in full-Stokes Zeeman Doppler Imaging techniques (e.g., \citeads{2015ApJ...805..169R}).}, our proximity to the Sun provides a high-resolution insight into the outer layers of its atmosphere. The solar atmosphere poses a wonderful laboratory to study stellar atmospheres and the processes that are responsible for energy transport, and \emph{forward modeling} and \emph{inversion} have been particularly successful for understanding line formation and in the interpretation of spectropolarimetric and spectroscopic observations.

Forward modeling techniques usually involve magnetohydrodynamical simulations of a stellar atmosphere that may include waves, magnetic fields, flows and a number of dynamic events. Synthetic spectra from those model atmospheres can be qualitatively compared with observations to study plasmas or to understand how spectral lines react/form under different physical conditions. Recent developments in 3D MHD simulations have made calculations that simultaneously include a photosphere, a chromosphere and a corona possible, and these have been extensively used to study the elusive solar interface region.

Inversion techniques rely upon algorithms that systematically modify the parameters of a \emph{guessed} atmospheric model, to minimize the difference between the emerging spectra from that model and the observations. Therefore, inversions can infer quantitative information from an observation, assuming that the proposed model and radiative transfer can \emph{realistically} describe the conditions in the atmosphere. Inversions have been extensively used to analyze photospheric observations and, more recently, to study the chromosphere. 

\noindent In this chapter we review the state-of-the-art of radiative diagnostics in solar physics. In \S\ref{sec:rad} we introduce radiative transfer and response functions. In \S\ref{sec:mod} and \S\ref{sec:diagno} we describe inversion techniques and the main spectral lines used in chromospheric studies. In \S\ref{sec:chrodiag}, some recent inversion results are presented. Finally, future perspectives of radiative diagnostics are discussed in \S\ref{sec:fut}.

%% file: response.tex
In the field of stellar and solar physics, almost all knowledge that has been accumulated has been acquired through the analysis and interpretation of radiation, combined with the known laws of physics. The field is thus crucially dependent on the correct understanding of the information contained in the light that we receive.

The field of radiative transfer has a rich history, dating back to the 19th century, when the study of stellar and solar light led to the discovery of many new elements, some of which, such as Helium, were even discovered first on the Sun, before they were found also on Earth. Particularly the first half of the 20th century saw a revolution in our understanding of the structure of atoms and molecules, and the interaction of light with them, which layed the foundations for the inferrence of the composition and conditions of stellar atmospheres.

\subsection{Radiative transfer in stellar atmospheres}
In stellar atmospheres, the structure of the atmosphere is dominated by the transport of the energy, generated in the core of the star, to the outside. In the part of the atmosphere that is visible to an observer, the conversion of thermal energy to radiation is thus of crucial importance for the vertical structure, and more importantly is, at least in part, determined by the radiation field itself. The radiation thus carries the only information we receive from any stellar atmosphere, and at the same time is an essential constituent of that same atmosphere.

For the correct interpretation of the light that we receive, it is thus necessary to self-consistently model the radiation field and the atmospheric structure, by solving the radiative transfer equation and the equations of statistical equilibrium, along with hydrostatic equilibrium. 

The monochromatic radiative transfer equation (RTE) for a single ray of polarized light can be expressed as
\begin{equation}
 \mu \frac{d\mathbi{I}}{ds}  = -\mathbf{K}\mathbi{I}+ \mathbi{j},\label{eq:rt}
\end{equation}
where $\mathbi{I}=(I,Q,U,V)^T$ is the Stokes parameter vector, 
$\mathbi{j}=(j_I,j_Q,j_U,j_V)^T$ \ is
the total emission vector, \kv \ is the total absorption matrix, and $\mu$ is the angle of the ray relative to the surface.
There are seven independent terms in the absorption matrix: $k_I$ is related to the line opacity; $k_Q$,
$k_U$ and $k_V$ describe the coupling of the intensity $I$
with $Q$, $U$ and $V$; and $f_Q$, $f_U$ and $f_V$ are
conversion terms between $Q$, $U$ and $V$ due to magneto-optical
effects (\citeads{2004ASSL..307.....L}):
\begin{equation}
 \mathbf{K} = 
 k_c \begin{pmatrix}
      1 \ & 0 \ & 0\ & 0 \  \\[0.3em]
      0 & 1 & 0 & 0  \\[0.3em]
      0 & 0 & 1 & 0  \\[0.3em]
      0 & 0 & 0 & 1  \\[0.3em]
      \end{pmatrix} + k_l  \begin{pmatrix}
      k_I & k_Q & k_U & k_V \\[0.3em]
      k_Q & k_I & f_V & -f_U \\[0.3em]
      k_U & -f_V & k_I & f_Q \\[0.3em]
      k_V & f_U & -f_Q & k_I \\[0.3em]
      \end{pmatrix}.\label{eq:abmat}
\end{equation}
where $k_c$ is the background opacity and $k_l$ is the line opacity. In general, the line opacity depends on the populations of the lower and upper level of the transition ($n_l$ and $n_u$), the oscillator strength of the transition ($f$), the statistical weight of the atomic levels ($g_l$ and $g_u$) and the Doppler width of the line ($\Delta\nu_D$),
\begin{equation*}
k_l = \frac{\pi e_0^2}{mc} \frac{n_l f}{\Delta \nu_D}\bigg (1 - \frac{n_u g_l}{n_l g_u} \bigg ),
\end{equation*}
where the Doppler width of the line is given by
\begin{equation}
\Delta\nu_D = \frac{\nu_0}{c}\sqrt{\frac{2TK_B }{M} + \mathrm{v}_{turb}^2}, 
\end{equation}
where $K_B$ is the Boltzmann constant, $T$ is the temperature, $M$ is the atomic mass and $\mathrm{v}_{turb}$ is the microturbulent Doppler broadening.

The terms involved in the RTE are normally derived from the physical parameters of the model atmosphere: temperature, macroscopic line-of-sight velocity, turbulent motions and the magnetic field vector. In practice, the parameters of the model atmosphere are usually known on a discrete grid of depth points along the ray. In most cases, the depth-stratification of  $\mathbf{K}$ and $\mathbi{j}$ are not known analytically and the RTE must be integrated numerically assuming some boundary condition. Over the years, many methods have been developed to integrate the RTE to obtain the emerging full-Stokes intensity. In the early 70s, it was common to use a Runge-Kutta solver (\citeads{1976A&AS...25..379L}), but this type of solver is very inefficient and shows poor accuracy on coarse depth grids. The \emph{diagonal element lambda operator} method (DELO, \citeads{1989ApJ...339.1093R}) reformulates the RTE in terms of the \emph{source vector} $(\mathbi{S} = \mathbf{j}/k_I)^T$ by dividing all terms in Eq.~(\ref{eq:rt}) by the absoption coefficient $k_I$. This change leaves all diagonal terms in $\mathbf{K}$ equal to unity, which allows for the substitution
\begin{equation*}
\mathbf{\bar{K}}=\frac{\mathbf{K}}{\eta_I} - \mathbbm{1},
\end{equation*}
where $\mathbbm{1}$ is the $4\times4$ identity matrix. The RTE becomes:
\begin{equation}
  \mu \frac{d\mathbi{I}}{d \tau} = \mathbi{I} - \mathscr{S},\label{eq:DELO}
\end{equation}
with the monochromatic optical depth defined as $d\tau = k_I ds$ and $\mathscr{S} = \mathbi{S} - \mathbf{\bar{K}}\mathbi{I}$. 
On a discrete grid of depth points the solution is normally expressed as an integration starting at the \emph{upwind} point, towards the next point in the direction of propagation (usually referred as the \emph{central} point), which we denote with the sub-indexes $u$ and $0$ respectively: 
\begin{equation*}
  \mathbi{I}(\tau_0) = \mathbi{I}(\tau_{u}) \mathbf{O}(\delta\tau_{0})+ \int_{\tau_{u}}^{\tau_{0}} \mathbf{O}(t)\mathscr{S}(t)dt.
  \end{equation*}
$\mathbf{O}$ is the evolution operator, a $4\times4$ real matrix that applied to the Stokes vector at point $u$ yields the Stokes vector at point $0$. The evolution operator can have a complicated analytical form, but assuming that the absorption matrix is constant between two consecutive grid points and equal to the value in the middle of the interval, the evolution operator can be expressed in terms of exponential terms. \begin{equation}
  \mathbi{I}(\tau_0) = \mathbi{I}(\tau_{u}) e^{-\delta\tau_0} + \int_{\tau_{u}}^{\tau_{0}} e^{-(t-\tau_{u})}\mathscr{S}(t)dt.\label{eq:DELOsol}
\end{equation}
{An underlying assumption of the DELO method is that the diagonal terms of $\mathbf{{K}}$ are much smaller than the non-diagonal terms that are included in the effective source vector $\mathscr{S}$. This way of writing the transfer equation has allowed us to use the same evolution operator as in the unpolarized case, but we have conveniently hidden some dirt under the effective source vector. Strictly speaking the evolution operator is exact only when the non-diagonal terms of $\mathbf{{K}}$ are zero, but this approximation seems to hold rather well when compared against other methods for solving the PRT equation.}

Eq.~\ref{eq:DELOsol} can be integrated \emph{analytically} if the equivalent source vector is assumed to have a given dependence with depth within the interval $(\tau_u, \tau_0)$. The solution can be written as a function of the ensuing intensity and the upwind, center and (sometimes) downwind values of the modified source function. \citetads{1989ApJ...339.1093R} assumed a linear dependence giving rise to the DELO-Linear solution. \citetads{2003ASPC..288..551T} showed that the linear approximation used in the DELO-Linear yields a poor absolute error in the computation of the atomic populations and developed the DELO-Parabolic method by splitting the equivalent source vector in the integral term of Eq.~(\ref{eq:DELOsol}) and assuming a parabolic dependence for $\mathbi{S}$ and linear dependence for $\mathbf{\bar{K}}\mathbi{I}$. 
A problem of the parabolic dependence is that parabolae can easily overshoot if the monochromatic depth scale is not equidistant. Therefore, \citetads{2013ApJ...764...33D} introduced two integration strategies based on non-overshooting Bezier splines that solve the issues of DELO-parabolic while keeping quadratic and cubic accuracy, usually referred as DELO-Bezier. The DELO solvers can be computed with relatively low computational cost and usually provide good accuracy in most situations, making them especially valuable so compute non-LTE problems and inversions where the RTE must be solved accurately many times. An alternative Hermitian method proposed by \citetads{1998ApJ...506..805B} has also been extensively used in solar inversion codes. 

{A different approach to solve the PRT equation has been proposed by \citetads{2016A&A...586A..42S}. Instead of searching for a smooth interpolant, they allow for a discontinuous solution at the boundary of each grid cell in the atmosphere. Within each cell, the source function is allowed to change linearly and the PRT equation is integrated using the Milne-Eddington solution. This approach shows very promising results, and is especially accurate when coarse grids are used to describe the model atmosphere, a property that is especially relevant for inversion methods. However, the convergence stability of this approach when used to solve the statistical equilibrium equations with approximate lambda operators remains to be explored.}

\subsection{Coupling between atmospheric and radiative quantities}\label{sec:rtterms}
We now consider the dependencies of the RTE on the parameters of the model atmosphere.  Assuming a known set of physical quantities, such as temperature, line of sight velocity, turbulent velocities, the three components of the magnetic field vector, and a depth scale, it is possible to compute the radiative properties that appear in Eq.~(\ref{eq:DELO}).  For reasons of simplicity, we first consider here the case of non-polarized radiation, in which case the radiative transfer equation can be written as:
\begin{equation}
\mu \frac{dI_{\mu\nu}}{ds} = \eta_{\mu\nu} - \chi_{\mu\nu} I_{\mu\nu},\label{eq:nopol}
\end{equation}
where $\eta_{\mu\nu}$  and $\chi_{\mu\nu}$ are the total emissivity and opacity, including continuum and line contributions. Eq.~(\ref{eq:nopol}) can be re-written using the source function $S_{\mu\nu} = \eta_{\mu\nu} / \chi_{\mu\nu}$ as
\begin{equation*}
\mu \frac{dI_{\mu\nu}}{-\chi_{\mu\nu} ds} =  I_{\mu\nu} - S_{\mu\nu}  \Rightarrow  \mu \frac{dI_{\mu\nu}}{d\tau_{\mu\nu}} =  I_{\mu\nu} - S_{\mu\nu},
\end{equation*}
where $d\tau_{\mu\nu} = -\chi_{\mu\nu} ds$ is the optical depth along the ray. 

The line emissivity ($\eta_\nu$) and opacity ($\chi_\nu$) characterize the line radiative properties as a function of the atomic level populations $n$ and the Einstein coefficients ($B_{ll'}$, $B_{l'l}$ and $A_{ll'}$). In the unpolarized case and assuming complete redistribution (CRD) of scattered photons, the expressions are simplified to the following when the energy of level $l'$ is smaller than that of level $l$ ($l\succ l'$, using the same notation as \citeads{1991A&A...245..171R}):
\begin{eqnarray}
\eta_{ll'}(\mu,\nu) = \frac{h\nu}{4\pi}n_u A_{ll'} \varphi(\mu, \nu),\label{eq:emi}\\
\chi_{ll'}(\mu,\nu) = \frac{h\nu}{4\pi} (n_l B_{l'l} - n_u B_{ll'}) \varphi(\mu, \nu),\label{eq:opac}
\end{eqnarray}
where $\varphi$ is the profile, which we have imposed to be the same when a photon is absorbed or emitted (CRD). The total emissivity and opacity can be expressed as the sum of all contributions:
\begin{eqnarray*}
	\eta_{\mu\nu} = \sum_{l \succ l'} \eta_{ll'}(\mu,\nu) + \eta_c(\nu)\\
	\chi_{\mu\nu} = \sum_{l \succ l'} \chi_{ll'}(\mu,\nu) + \chi_c(\nu)
\end{eqnarray*}
where $\eta_c(\nu)$ and $\chi_c(\nu)$ are the continuum contribution to the total emissivity and opacity. For  time-independent equilibrium situations, the populations of the atomic levels $n$ can be obtained by solving the statistical equilibrium equations:
\begin{equation}
  \begin{split}
\sum_{l'\prec l} \bigg [ n_l A_{ll'} - (n_{l'} B_{l'l} - n_l B_{ll'}) \bar{J} \bigg ] - \\
\sum_{l' \succ l} \bigg [n_{l'} A_{l'l} - (n_l B_{ll'} - n_{l'} B_{l'l}) \bar{J} \bigg ]= \\- 
\sum_{l'} (n_l C_{ll'} + n_{l'}C_{l'l}) 
\label{eq:stateq}
\end{split}
\end{equation}
where $C_{ll'}$ is the collisional rate between level $l$ and $l'$, which is generally a function of temperature and density only, and $\hat{J}$ is the integrated mean intensity (CRD) given by
\begin{equation*}
\bar{J} = \frac{1}{4\pi} \int d\Omega \int d\nu \varphi(\mu,\nu) I_{\mu,\nu},
\end{equation*}
which depends on temperature, density and LOS velocity, as well as any microturbulent velocity.

In the unpolarized case then, the dependence of all elements in Eq.~(\ref{eq:stateq})  on the atmospheric quantities in most cases not especially complicated. However, calculating the radiative quantities $\alpha$ and $\eta$ for a given atmosphere is by no means a straightforward task. The principal difficulty in solving (\ref{eq:stateq}) is that we need to know the atomic level populations to calculate the intensities using Eq.~(\ref{eq:emi}) and (\ref{eq:opac}), but the atomic populations depend on the intensity through $\bar{J}$. Therefore, the problem is not only non-local, it also highly non-linear and must therefore be solved iteratively, using an appropriate approximation to describe the dependence of  $\bar{J}$ on the the atomic level populations, as pioneered by \citetads{1973JQSRT..13..627C}, \citetads{1981ApJ...249..720S}, \citetads{1987JQSRT..38..325O} and \citetads{1991A&A...245..171R} among others.  For every set of temperature, density and velocity, however, a solution can be calculated.

In the general polarized case,  Eq.~(\ref{eq:stateq}) contains the population densities of all substates of every energy level, which increases both the size and the difficulty of calculating the coefficients of the linear system to solve significantly. More specifically, the energy of the various sub-states within each energy level is now dependent on the magnetic field strength and direction. Although this is what makes it possible to quantify the magnetic field strength and direction in the solar atmosphere, the associated computational cost is significant. Many synthesis codes therefore resort to the use of the \emph{polarization-free} approximation, that first solves for the populations using the unpolarized approximation, and  distributes the populations of the unpolarized levels across the sub-levels according to their statistical weight, thus allowing for a single full-Stokes solution to be computed  (\citeads{1996SoPh..164..135T}).

Although the polarization-free approximation offers a great simplification, in many cases the density is sufficiently high that the collisional rates dominate  Eq.~(\ref{eq:stateq}), and the solution can be accurately described by the approximation of LTE. Under these conditions, the source function is directly given by the Planck function, and in practice we only need to compute explicitly the terms in the absorption matrix. Assuming that only the Zeeman effect is at work hereafter, the terms of the absorption matrix in Eq.~(\ref{eq:abmat}) are given by the following expressions (\citeads{2004ASSL..307.....L}):
\begin{eqnarray*}
k_I &=& \frac{1}{2}\bigg [ h_p \sin^2\theta + \frac{h_b + h_r}{2}(1+\cos^2\theta) \bigg ]\\
k_Q &=& \frac{1}{2}\bigg [h_p -  \frac{h_b + h_r}{2} \bigg ] \sin^2\theta \cos 2\chi\\
k_U &=& \frac{1}{2}\bigg [h_p - \frac{h_b + h_r}{2} \bigg ]\sin^2\theta \sin 2\chi\\
k_V &=& \frac{1}{2}\bigg [h_r - h_b   \bigg ]\cos\theta\\
f_Q &=& \frac{1}{2}\bigg [g_p -  \frac{g_b + g_r}{2}     \bigg ] \sin^2\theta \cos 2\chi\\
f_U &=& \frac{1}{2}\bigg [g_p  - \frac{g_b + g_r}{2}   \bigg ] \sin^2\theta \sin 2\chi\\
f_V &=& \frac{1}{2}\bigg [g_r - g_b \bigg ] \cos\theta
\end{eqnarray*}
where $\theta$ and $\chi$ are the inclination and the azimuth of the magnetic field vector and the $h$ and $g$ components are computed from the Voigt-Faraday functions including the effect of Doppler motions and the Zeeman splitting. The latter is normally characterized by a wavelength shift of the polarized components and their relative strength.

The wavelength shift of the Zeeman components ($\Delta\lambda_B$) can be expressed in terms of the quantum state $M$ according to the selection rules:
\begin{equation*}
	\Delta M = M_u - M_l = \left\{\begin{matrix}
 +1 &\equiv & b \\ 
 0 &\equiv & p \\
 -1 &\equiv & r 
\end{matrix}\right. 
\end{equation*}
The computation of the profiles is directly given by the following expressions (for $q=b,p,r$):
\begin{eqnarray*}
h_q = \sum_{M_l M_u} S_q^{J_lJ_u}\frac{1}{\Delta\nu_D \sqrt{\pi}} H(a, v-v_a + v_B(g_uM_u - g_lM_l))\\
f_q = \sum_{M_l M_u} S_q^{J_lJ_u}\frac{1}{\Delta\nu_D \sqrt{\pi}} L(a, v-v_a + v_B(g_uM_u - g_lM_l)),
\end{eqnarray*}
where $S_q$ is the strength of the Zeeman component, $H$ and $L$ are the {absorption and dispersion profiles} respectively, $a$ is the damping and $v$, $v_a$ and $v_B$ are the frequency, Doppler velocity ($V_{l.o.s.}$) and Zeeman splitting respectively expressed in Doppler width units,
\begin{equation}
  \begin{split}
  v = \frac{\nu_0 - \nu}{\Delta\nu_D}, \ \ v_a = \frac{\nu_0 V_{l.o.s}}{c\Delta\nu_D}, \ \ v_B =  \frac{e_0}{4\pi m_e c}\frac{B}{\Delta\nu_D}, \\
  a = \frac{\Gamma_{stark} + \Gamma_{vdW} + \Gamma_{rad}}{\Delta\nu_D}.
  \end{split}
\end{equation}
We have chosen to work with frequencies instead of wavelength, but $v_B$ can also be expressed in terms of the central wavelength of the line $\lambda_0$ (Eq.~\ref{eq:vb}). This form of $v_B$ explicitly shows that the Zeeman splitting scales quadratically with $\lambda$ and linearly with $B$ (with $\Delta\lambda_D$ the Doppler width in units of wavelength),
\begin{equation}
 v_B = \frac{e_0}{4 \pi m_e c }\frac{ B \lambda_0^2}{\Delta \lambda_D}.\label{eq:vb}
\end{equation}

The damping parameters originate from collisions with charged particles ($\Gamma_{stark}$), collisions with neutral particles ($\Gamma_{vdW}$) and from the finite lifetime of atomic levels  ($\Gamma_{rad}$). These parameters have traditionally been computed theoretically or measured and they are typically considered part of the input atomic data that can be retrieved from an atomic database like VALD (\citeads{1995A&AS..112..525P}; \citeads{1998PASA...15..336B}). The damping parameters are typically functions of the electron density, the hydrogen and helium densities and the temperature.

Finally, the strength of the Zeeman components can be computed in general using $3-j$ symbols ($S_q^{J_l J_u})$, but in the case of the Zeeman effect they are given in Table~\ref{tab:zeeman} as a function of the $J$ and $M$ quantum numbers of the sub-levels of the transition.
\begin{table*}[!ht]
\centering
\begin{tabular}{ >{$}c<{$} | >{$}c<{$} | >{$}c<{$} | >{$}c<{$} |}
 \cline{2-4} & J_u = J_l + 1 & J_u = J_l & J_u = J_l - 1 \TBstrut\\
 \hline
 \multicolumn{1}{|>{$}c<{$}|}{\Delta M = +1 \equiv b} & \displaystyle\frac{3(J_l + M_l +1)(J_l+M_l+2)}{2(J_l+1)(2J_l+1) (2J_l+3)} & \displaystyle\frac{3(J_l-M_l)(J_l+M_l+1)}{2J_l(J_l+1)(2J_l+1)} & \displaystyle\frac{3(J_l-M_l)(J_l-M_l-1)}{2J_l(2J_l-1)(2J_l+1)}\TBstrut\\
 \hline
 \multicolumn{1}{|>{$}c<{$}|}{\Delta M = 0  \equiv p}  & \displaystyle\frac{3(J_l - M_l +1)(J_l+M_l+1)}{(J_l+1)(2J_l+1) (2J_l+3)} & \displaystyle\frac{3M_l^2}{J_l(J_l+1)(2J_l+1)} &  \displaystyle\frac{3(J_l-M_l)(J_l+M_l)}{J_l(2J_l-1)(2J_l+1)} \TBstrut\\
  \hline
 \multicolumn{1}{|>{$}c<{$}|}{\Delta M = - 1 \equiv r}& \displaystyle\frac{3(J_l - M_l +1)(J_l-M_l+2)}{2(J_l+1)(2J_l+1) (2J_l+3)} & \displaystyle\frac{3(J_l+M_l)(J_l-M_l+1)}{2J_l(J_l+1)(2J_l+1)} &  \displaystyle\frac{3(J_l+M_l)(J_l+M_l-1)}{2J_l(2J_l-1)(2J_l+1)} \TBstrut\\
 \hline
\end{tabular}
\caption{Strength of the Zeeman components as a function of the quantum numbers $J$ and $M$. For clarity, we note that $\Delta M = M_u - M_l$. Source: \citeads{2004ASSL..307.....L}, Table 3.1.}\label{tab:zeeman}
\end{table*}

Although the above outline is far from complete, it is possible to conclude that there is a limited number of atmospheric quantities that control the radiative properties of the solar atmosphere. While some of the dependencies on these quantities can be approximated in a fairly simple way, others enter the expressions in many different places, some of them both directly and indirectly. Perhaps the most convoluted dependency is that on the temperature, which appears in the expressions for the Doppler width, the line opacity, the source function and usually the damping constants. The densities and the ionization balance mostly influence the damping and the line opacity. The line of sight velocities and the magnetic field influence the absorption matrix, which encodes the Doppler shifts and Zeeman splitting at each depth of the atmospheric model.  In \S\ref{ssec:rf} we formalize and illustrate these dependencies using so-called \emph{response functions}.

\subsection{Response functions}\label{ssec:rf}
\begin{figure*}[hbt]
\centering
\includegraphics[width=0.8\textwidth, trim=0 0 0 0, clip]{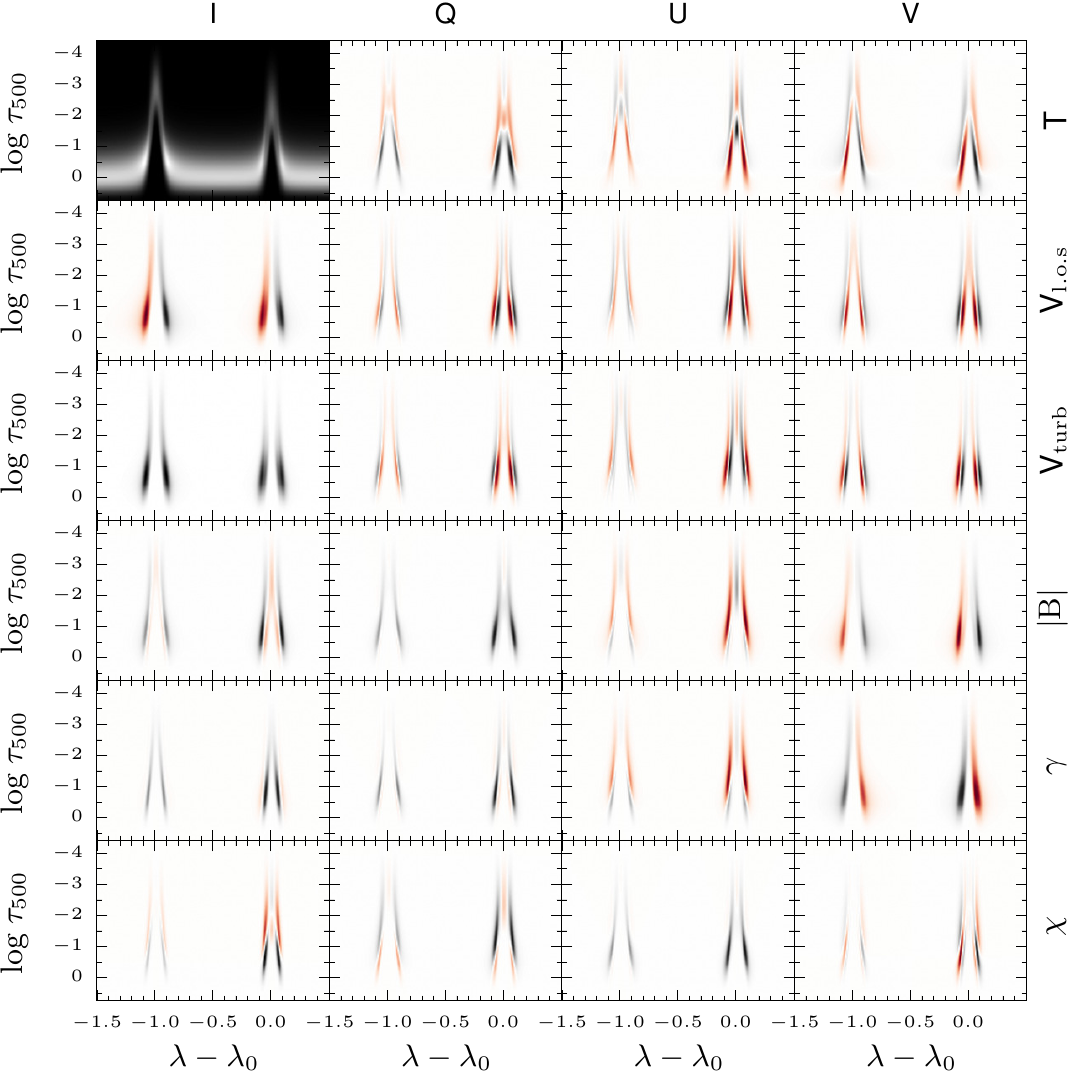}
\caption{Numerical response functions of Stokes~$I$, $Q$, $U$ and $V$ (from left to right) to the temperature, line of sight velocity, turbulent motions, magnetic field strength, magnetic field inclination ($\gamma$) and magnetic field azimuth ($\chi$) from top to bottom. The calculations have been performed in LTE using the FAL-C atmosphere with an adhoc magnetic field vector of 500 G.}\label{fig:rf}
\end{figure*}

In order to change a given atmosphere in such a way that the emerging spectrum matches an observed spectrum, information on how the spectrum changes when the atmosphere is perturbed must be obtained. Unfortunately, since the radiation influences the state of the atmosphere in a non-local and non-linear way, the response of the spectrum to the atmospheric perturbation is also not linear, so that a perfect correction generally cannot be produced at once. {In most cases the response}, although not linear, will reduce the difference, so that an iterative approach can be used. 

The response of the spectrum to an atmospheric perturbation can be calculated \emph{numerically} using finite differences, by simply calculating the spectrum, perturbing the atmosphere and calculating the spectrum again. The difference is then divided by the perturbation, yielding the desired response function. Alternatively, a \emph{centered} derivative (in the nominal value of the parameter) is likely to be more accurate but it will be twice as slow:
\begin{equation}
  \begin{split}
    R_\lambda(x,\tau_k)  \approx \underbrace{\frac{S_\lambda(x+\delta x, \tau_k) - S_\lambda(x-\delta x, \tau_k) }{2\delta x \Delta \tau_k}}_\mathrm{centered} \approx \\
    \underbrace{\frac{S_\lambda(x+\delta x, \tau_k) - S_\lambda(x, \tau_k) }{\delta x \Delta \tau_k}}_\mathrm{non-centered},
    \end{split}
\end{equation}
where $R_\lambda$ is the monochromatic response function to perturbations to a parameter $x$ and $S_\lambda$ is the monochromatic emerging Stokes vector. \emph{Response functions} for polarized light were proposed by \citetads{1977A&A....56..111L}, who derived an analytical expression by perturbing the RTE. However, those early studies did not couple directly the response of the spectra to changes in typical parameters of model atmospheres like, e.g., temperature or density but rather to perturbations of the parameters of the RTE. For depth-stratified atmospheres, analytical response functions have been derived in the LTE case (\citeads{1992ApJ...398..375R}), but it is not trivial to write them as the response of all parameters must be also be propagated through all the calculations involved in the computation of the terms of the RTE and into the equation of state of the gas. Analytical response function in Milne-Eddington codes are also commonly used and somewhat less convoluted to derive (e.g., \citeads{2007A&A...462.1137O}). The advantage of analytical response functions is that they are usually more precise and much more efficient to compute than the finite differences approach, as the overhead compared to synthesizing a single profile is small.

Fig.~\ref{fig:rf} illustrates the numerical response functions of the \ion{Fe}{i} $\lambda 6301.5$ and $\lambda 6302.5$ lines to temperature, line of sight velocity, turbulent broadening, magnetic field strength, inclination and azimuth along the optical depth scale of the FAL-C model atmosphere, including a constant magnetic field strength of $500$~G inclined at $45$~$\deg$\ with and azimuth of $20$~$\deg$.  It is beyond the scope of this review to analyze in detail each of the panels, since response functions are model dependent, but we note some basic properties that can be illustrated using these panels:
\begin{itemize}
\item Spectral lines sample a range of optical depths, in this case increasing in height from the continuum towards line center.
\item Of the 6 considered parameters, only the temperature influences the continuum.
\item The coupling effect of magneto-optical terms in the absorption matrix is clearly visible for example in the response of all Stokes parameters to changes in the azimuth (\citeads{1979SoPh...63..237L}).
\item In most parameters the shape of the response function in Stokes $Q$, $U$ \& $V$ can appear to be quite convoluted as an effect of the shape of the profiles.
\end{itemize}

%% file: model.tex
Unlike what is suggested by the name, most inversion codes do not provide an inverse mapping from the spectra to an atmosphere, but instead {\em infer} the atmosphere from the data in a semi-automated way. Since the mapping from atmosphere to spectrum is not reversible in general, simplifying assumptions generally need to be made to ensure the solution remains physically plausible. These assumptions have a significant impact on the reliability of the result, in a way that is not evident from either the precision with which the data are reproduced, nor from the physical plausibility of the inverted result. 

The restrictions imposed are often ad-hoc, because the amount of constrainable information in the spectrum is generally difficult to estimate, but sometimes some a-priori restrictions can be applied, based on physical arguments, that can add stability, allowing for less restrictive assumptions to be made elsewhere.

The basic ingredients of any inversion code are the ability to generate a synthetic spectrum through a forward calculation on an atmospheric model, knowledge of the important instrumental effects involved in mapping the synthetic spectrum to the observed data, and an automated minimization engine, which can be based on a number of techniques. 

\subsection{Atmospheric model}

Most inversion codes are fundamentally based on a physical formation model of a varying degree of complexity. There are two main areas in which the model can vary in complexity: the atmospheric structure and the physics assumed to be involved in generating the emerging spectrum.

\subsubsection{Model atmospheres}
The sophistication of the atmosphere that can be fitted to an observation depends to a large extent on the number and type of spectral lines that it is required to reproduce the profile of. Where a single spectral line contains already a large amount of information about the atmosphere in which it originated, the exact shape of several different line profiles can significantly restrict the possibilities.

For the inversion of a single spectral line, a commonly made choice is that of a Milne-Eddington (ME) atmosphere (\citealt{harvey1972line}; \citeads{1977SoPh...55...47A}), which is characterized by a constant opacity, linear source function, and a host of arbitrary fit parameters that are needed to fit the profile, but do not really correspond to useful physical information. The solution can be integrated analytically, however, leading to a very efficient calculation of the line profile, which allows for a fast inversion. This type of atmosphere is able to reproduce simple spectral line profiles. Many codes exist, that are based on this type of atmosphere (e.g.  HELIX \citepads{2004A&A...414.1109L}, MERLIN \citepads{1987ApJ...322..473S}, HMI VFISV \citepads{2011SoPh..273..267B}, to name just a few), each of them using different minimization strategies to find the best fit to the observed spectrum (see comparison by \citeads{2014A&A...572A..54B}). 

In those cases where more sophistication is needed to fit the observed spectra, either due to their large formation range, or due to their specific sensitivity to gradients in some physical quantity, a fully stratified, physical model of the atmosphere must be used. The advantage of this approach is that the fitted atmosphere has a firm physical basis, and can thus be used to study the magneto-hydrodynamic structure underlying an observation (\citeads{1992ApJ...398..375R}). 

\begin{figure*}[hbt]
\centering
 \includegraphics[width=0.8\textwidth,angle=0]{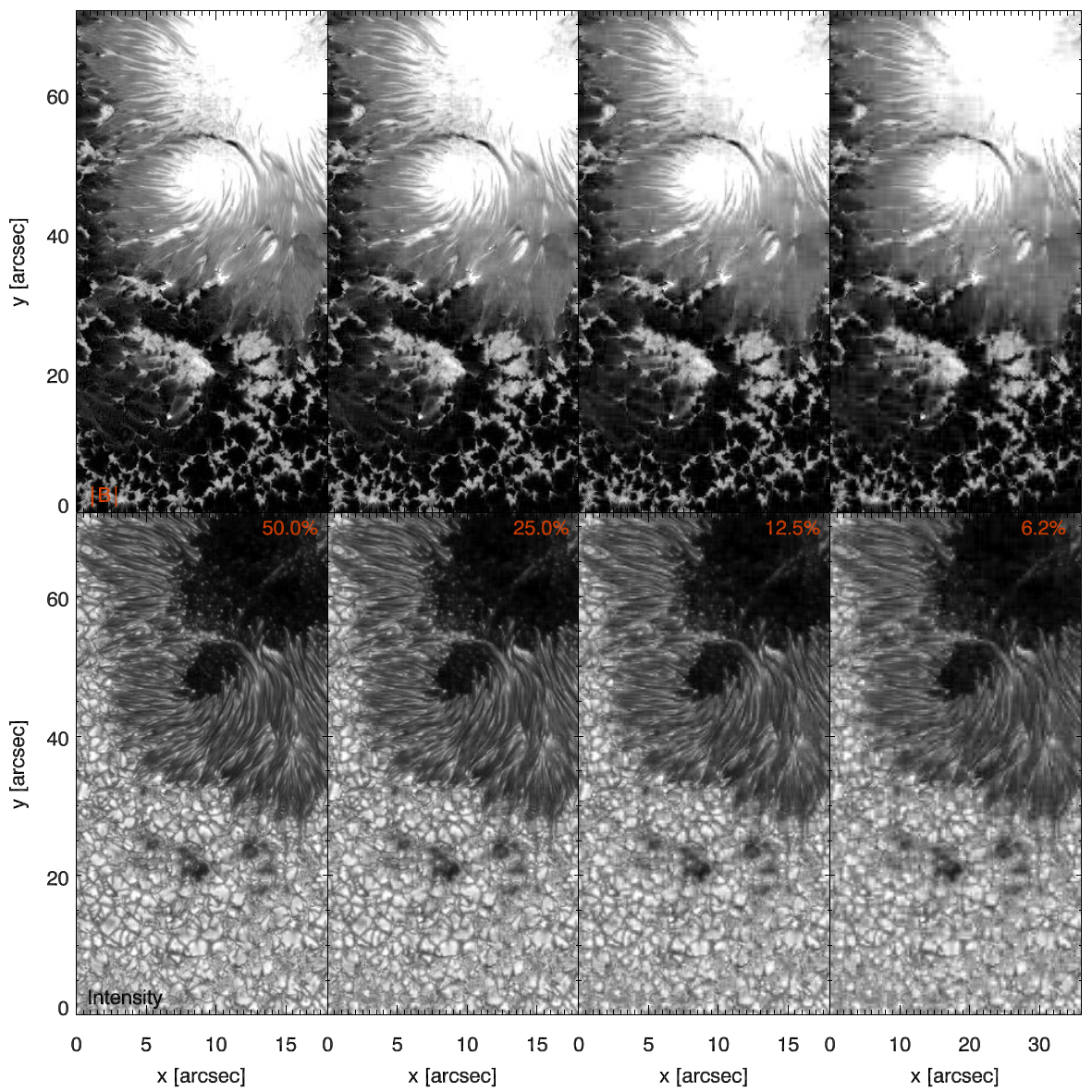}
  \caption{Compressibility of solar data from the inversion of a Hinode/SOT observation. \emph{Top:} Inferred magnetic field from a sparse inversion assuming a sparsity level of $50\%$ (left). The other maps have then been compressed assuming [25\%, 12.5\%, 6.25\%] sparsity. \emph{Bottom:} The emerging intensity in the wing of the \ion{Fe}{i}~$\lambda 6302$ line, computed from the sparse models above.}
  \label{fig:compress}
\end{figure*} 

A number of codes exists that allows for a height dependent stratification (SIR \citepads{1992ApJ...398..375R}, SPINOR \citepads{2000A&A...358.1109F}, NICOLE \citepads{1998ApJ...507..470S}, etc.), all of which make use of a gradient search based optimization method, using response functions or finite differencing to calculate the gradient. However, for all of them \emph{it is crucial that the level of stratification that is required to produce the spectrum is met, but not exceeded, or the inversion problem is no longer well posed.} This last problem in particular gives rise to a level of arbitrariness that is very difficult to quantify. 

More recently, a new generation of methods has begun development that tries to reduce some of the degrees of freedom of the inversions, by introducing additional constraints on the solution, that are either inspired by physics, or by other arguments. One of these is the sparse inversion code \citepads{2015A&A...577A.140A}, that requires the solution to be sparse, a quality that is inferred from the sparseness of the data (see Fig.~\ref{fig:compress}). This method is able to condense the information from many pixels in a reduced number of fit parameters, yielding a large reduction in the work needed to invert the data, and better constrained results.

\subsubsection{Physics}
The physics that is assumed to be of importance in the formation of the observed spectral lines can vary enormously, depending on the nature of the transition of interest, the spectral resolution, the signal to noise and the atmospheric heights of interest. Many inversion codes make specific assumptions about the atmosphere, but also about the physics believed to be important for the spectral line that was of primary interest to the author of the code at the time it was written. Here, we give a brief overview of the most important physical effects in which most of the major codes differ.

\begin{description}
\item[{\bf LTE/NLTE}]:

The most commonly employed assumptions are those of Local Thermodynamic Equilibrium (LTE) and plane-parallellity (1D). Both of these are never strictly applicable, since the mere observability of the radiation field implies the escape of energy and thus by definition non-local equilibrium conditions, and no object in the known universe is infinitely large and flat. Nonetheless, both approximations appear to be remarkably accurate in many cases of interest, and continue to enjoy popularity in the inversion community in particular, owing to the huge simplifications that can be made by using them.

The formation of many photospheric lines is well described using LTE, in which case the dimensionality of the problem can be disregarded altogether, and the spectrum can be directly and efficiently calculated for each image pixel from the magneto-hydrodynamic structure. For lines forming higher up in the solar atmosphere, however, this approximation often does not hold, and the assumption of LTE must be abandoned. Normally, this happens when collisional
rates are very low and radiation is weakly coupled
to the local physical conditions (e.g., in the chromosphere), invalidating the convenient assumption of Local Thermodynamical Equilibrium, that is commonly used in the
photosphere. This is particularly important for lines in which the scattering of radiation plays an important role, but comes at a hugely increased computational cost. Such costs must often be compensated for by simplifying the atmospheric model drastically, so that the overall sophistication of the formation model may not actually increase by much. 

\item[{\bf Geometry/Dimensionality}]:

Once the assumption of LTE is abandoned, the dimensionality of the atmosphere comes into play, since the non-locality of the radiation field manifests itself in the vertical, as well as in the horizontal dimensions. Although several forward solvers exist that can calculate the emergent spectrum in full 3D (e.g. RH, \citeads{2001ApJ...557..389U}; MULTI3D, \citeads{2009ASPC..415...87L}), no inversion code currently exists that can invert a dataset using full 3D radiative transfer, largely due to the huge computational effort this would require. 

\item[{\bf Hanle vs. Zeeman effect}]:

The calculation of the imprint that the magnetic field leaves in the emergent radiation, in the form of polarization, can depend on many details in the atomic physics, that influence the formation of the line profiles. Two different effects exist that can be used to extract the strength and orientation of the magnetic field in the line forming region: the Zeeman effect and the Hanle effect. While the Zeeman effect is the most widely used, owing to its simplicity and relatively high signal level, the Hanle effect can have a much higher sensitivity to weak magnetic fields, and it is an important tool for the characterization of magnetic fields in the low-density chromosphere.

The Zeeman effect \citepads{1897ApJ.....5..332Z} has been known for over a century and the theoretical description of it is relatively straightforward. Interpretation of Zeeman induced polarization has therefore been the cornerstone of photospheric inversions for many decades, in particular for the inversion of more complex, height-dependent atmospheres. However, its sensitivity to the magnetic field has intrinsic limitations: {in regions where the field strength is very weak and collisional rates are low, scattering polarization can dominate over Zeeman induced polarization. }

The use of the Hanle effect \citepads{1924ZfPh.....5..332Z}, is much less widespread, largely due to the complex physics and the large uncertainties in some of the physical quantities that are important for the interpretation, but it is able to return information on much weaker magnetic fields, especially in very low density regions. This effect has been the workhorse for measurements of the very weak magnetic field in chromospheric structures, such as loops and filaments.

\item[{\bf Molecules}]:

While most inversions are based on the interpretation of a single atomic transition, clearly the use of a larger number of transitions, where both effects are present, would constitute a much better constrained problem. Especially in cool regions, such as sunspot umbrae, the presence of molecular lines provides an opportunity to make use of very different line transition properties to constrain the atmospheric parameters. Significant progress has been made in that direction by a number of groups \citepads{2000A&A...364L.101B,2005ApJ...623L..57A}, but the complexity of the physics and the numerical cost have thus far remained substantial and few good inversions with molecules currently exist.

\end{description}
While this list is not intended to be exhaustive, it covers the most important differences in the physics between codes. 

\subsection{Data formation}
There are many ways in which an instrument can modify the true spectra while converting them to digital form. Any of them, when neglected, can alter the fitted atmospheric properties in various ways, which must be considered when interpreting the result.

\subsubsection{Spectral degradation}
\begin{figure*}[hbt]
\centering
\includegraphics[width=0.9\textwidth]{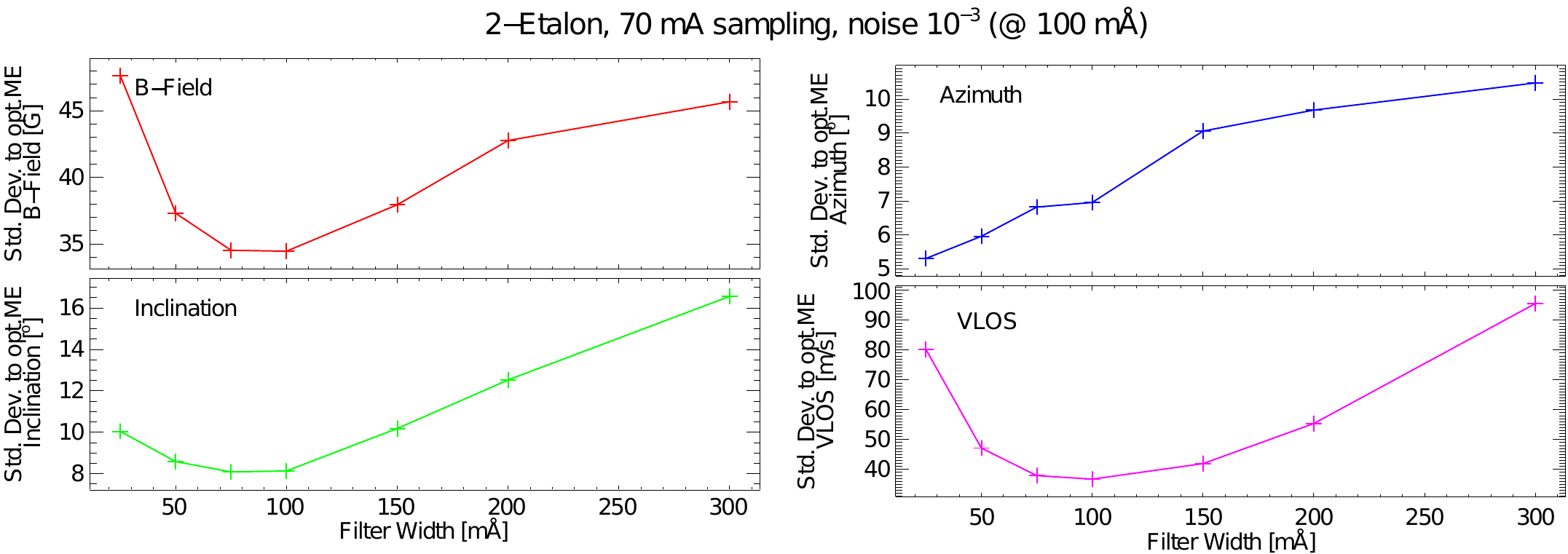}
\caption{Effect of spectral resolution in the accuracy of a Milne-Eddington inversion, using 5 spectral points. Courtesy: A. Lagg.}
\label{fig:sophires}
\end{figure*}
Traditionally, the formation of the observed data from a synthetic, undegraded spectrum, has involved mainly consideration of the spectral response of the instrument used to record the data. This is a simple effect to take into account, since it involves only the convolution of the synthetic spectra with the instrumental spectral response function, after which a direct comparison with the observed data can be made.

The spectral resolution of the observation is one of the most important factors in the ability of an inversion code to recover information from the spectral line, and determines to a large extent how complicated an atmosphere can be recovered from the data. There is such a direct link between the complexity of the atmosphere and the spectral resolution, that several instruments were designed and built, with the specific intent to produce data suitable for ME inversion codes, that have specifically optimized their spectral resolution to achieve the best result. Figure~\ref{fig:sophires} is an example of a study of this effect for the Solar Orbiter PHI instrument, and is the result of the inversion of sythesized observations. The error in the recovered inversion results decreases with increasing spectral resolution for almost all inverted quantities, but then increases again below a critical value, where aliasing of high-frequency spectral information starts to occur. This result suggests that since a ME inversion needs only 5 points to be adequately constrained, and the result is most accurate when the degraded spectrum is critically sampled, there is a maximum spectral resolution, beyond which the accuracy is not increased. {Obviously, better wavelength coverage and spectral resolution are usually desired, as long as the observations are  sampled accordingly. This example is particularly relevant for observations acquired with Fabry-Perot interferometers and Lyot filters, where wavelength coverage can be sacrificed in favor of temporal cadence for a given integration time.}

\begin{figure*}[hbt]
\centering
\includegraphics[width=0.9\textwidth]{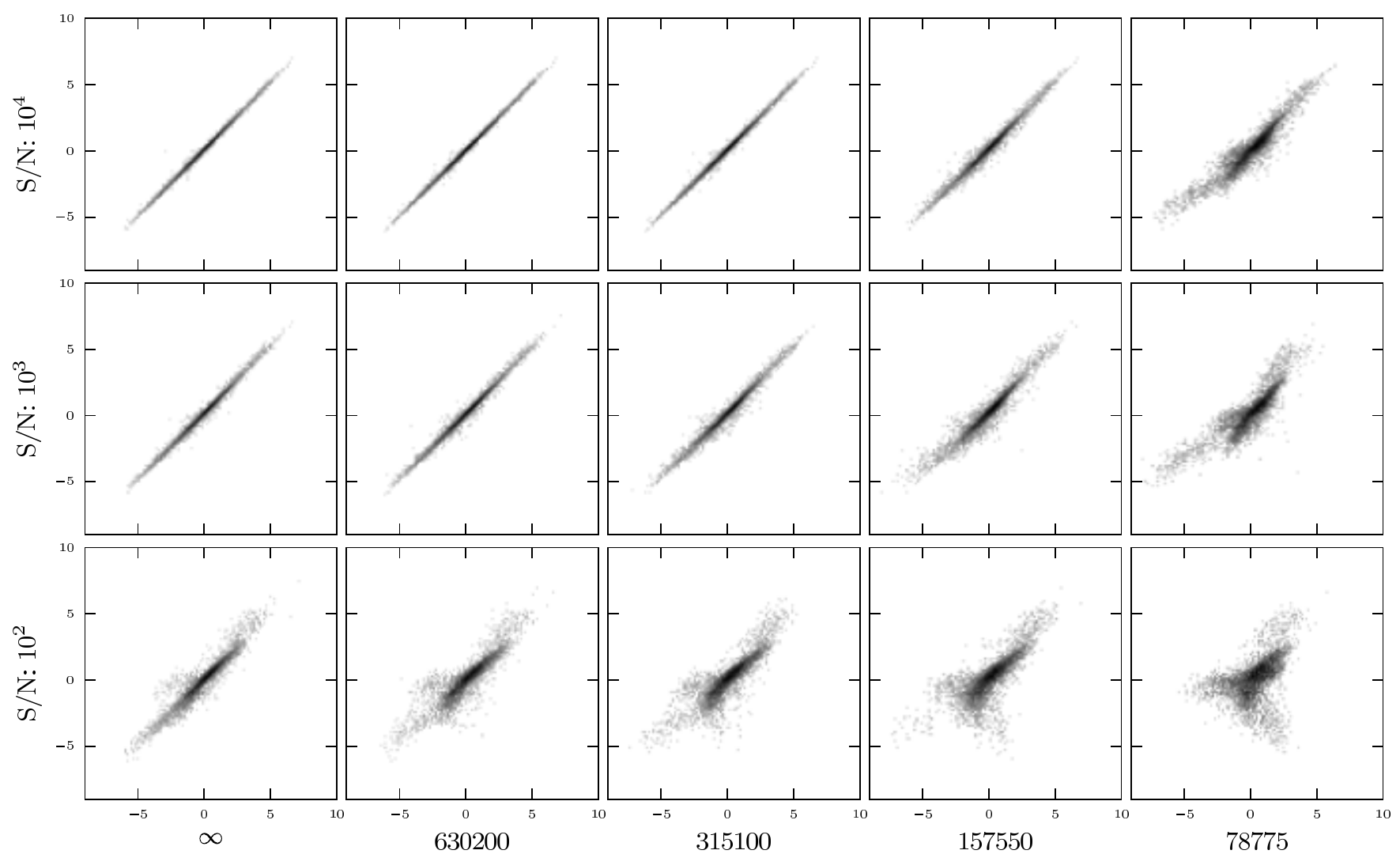}
\caption{Effect of spectral resolution vs. S/N on the recovered LOS velocity of a photospheric inversion based on simulated spectra of the FeI lines at 6301.5 and 6302.5\AA}
\label{fig:scatterplot_48_sd_cmp_vlos}
\end{figure*}

If a height dependent atmosphere is desired, however, the spectral resolution needs to be sufficiently high to obtain a reliable result. An illustration of this effect is given in Figure~\ref{fig:scatterplot_48_sd_cmp_vlos}, where the sythesized line profiles of the photospheric FeI lines at 6301.5 and 6302.5\AA\ of a quiet Sun MHD simulation, calculated using the MURaM code, were inverted using SPINOR assuming a height dependent atmospheric model, for several different amounts of noise and degraded using different spectral resolutions. The scatter, an indication of the accuracy with which the ``true'' value was recovered, although not negligible for noisless, full-resolution data, clearly increases with increasing noise level. It also increases rapidly with decrease in the spectral resolution, where a factor 2 in spectral resolution corresponds roughly to an order of magnitude in S/N. Most likely, the subtle variations in the line profile are reduced by a low spectral resolution, and are subsequently more easily overpowered by the noise.

\subsubsection{Spatial degradation}
Traditionally, the contamination of spectra with spectra from other parts of the field-of-view (FOV) has been considered as part of the formation process, most commonly in the form of a stray-light contribution of some sort, that is added to the undegraded spectra before or after degradation by the instrument. Typically, since the actual stray-light contribution depends on many factors, such as the observing conditions, time of day, etc., the stray-light profile is calculated from the average profile over the FOV or is modeled using a ``typical'' average quiet Sun atmosphere, and the amount of stray-light is fitted for as a free parameter (e.g. \citeads{1968AULII.....2..2S,1991sopo.work..307S,1996SoPh..164..277B}, and many more).

However, in most cases, the true nature of such {\em spatial} contamination is much more complicated, because there is a, more or less compact, mapping from the source (the undegraded spectra), to the data. This implies that to calculate the stray-light profile for a particular pixel, the undegraded spectra around it are required, which are not available, as they are themselves contaminated by an unknown amount of stray-light.

\citetads{2007PASJ...59S.837O} deal with this problem by using the data itself, and subtracting it from the original data by fitting. While this so-called ``local-straylight'' method indeed removes a lot of the unwanted stray-light, it has the significant disadvantage that it approximates a convolution of the real profiles with the convolution of the data, which has already been convolved. 

If spectra are available over an extended FOV, deconvolution can be considered to remove the spatial degradations directly. While it is in principle possible to recover the undegraded spectra by deconvolution, this procedure will amplify any noise in the data, preferentially at the highest possible spatial frequencies. One way to reduce this problem is to first accumulate the signal in the spectra in a reduced number of quantities, for instance by using a principal component analysis (PCA, \citeads{2000A&A...355..759R}), and then deconvolve the maps \citepads{2013A&A...549L...4R}. This method relies on the PCA basis functions of the data to be a good basis for the undegraded profiles, which for Hinode SP data seems to be a good approximation. Regularized spatial deconvolution has also been used to compensate the effect of straylight from atmospheric high-order aberrations (\citeads{2013A&A...553A..63S}).

\begin{figure*}[hbt]
   \centering
\includegraphics[width=0.75\textwidth,angle=0]{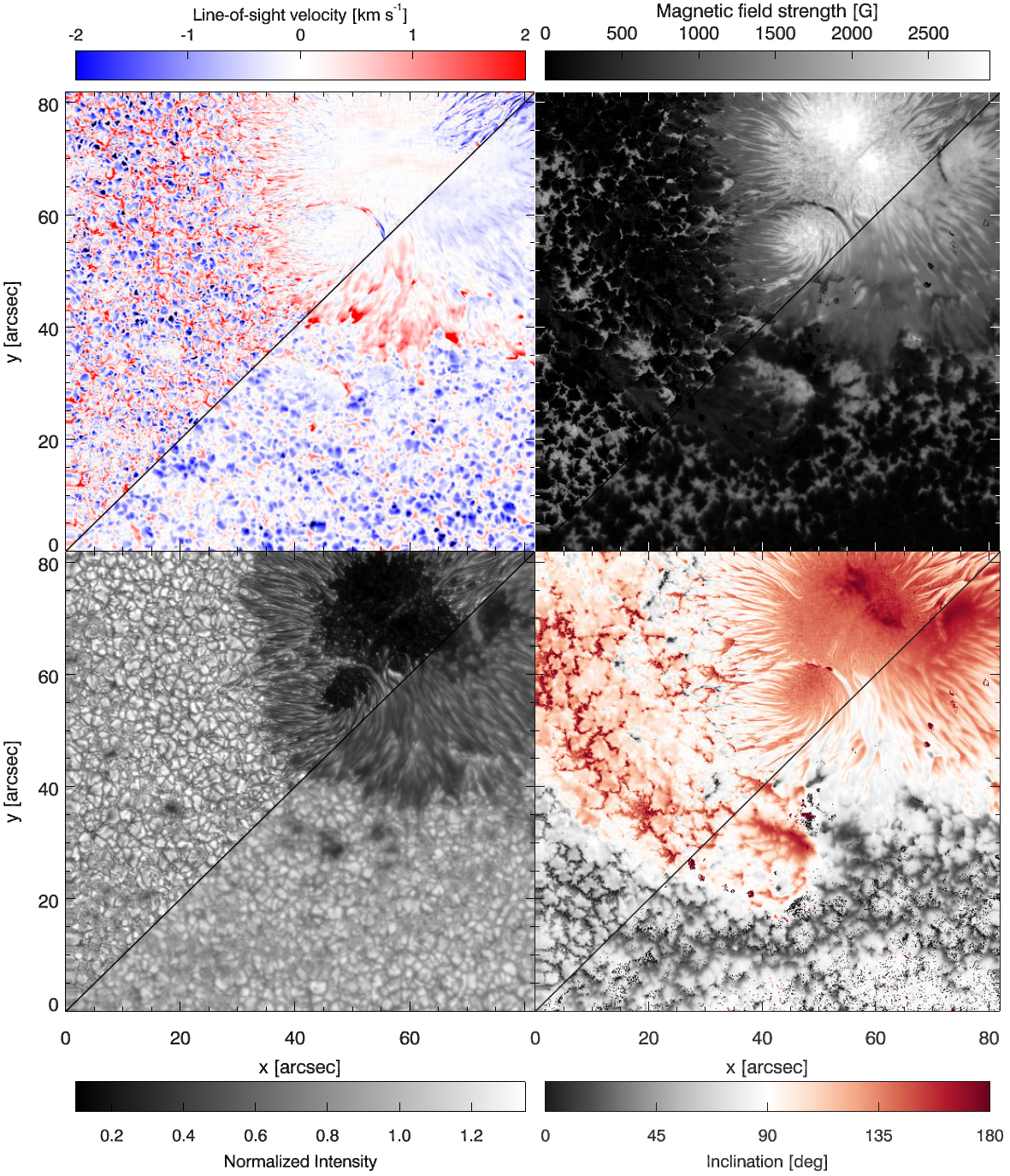}
  \caption{Coupled inversion of a sunspot (top left of each panel) observed with Hinode SOT/SP on the 5th of June 2007, compared with a "traditional" inversion, where stray-light is ignored (bottom right of each panel). Observation acquired with Hinode/SOT (\citeads{2007SoPh..243....3K}).}
  \label{fig:map}
\end{figure*} 
Another approach to calculate the stray-light is to use the actual undegraded profiles and apply the spatial degradation to them. Since the undegraded spectra are not know, we must self-consistently calculate the degraded and the non-degraded spectra, so that the line profiles calculated from the inversion, degraded by telescope diffraction, produces the observed data. This method is known as a coupled inversion \citepads{2012A&A...548A...5V}, and requires the simultaneous fitting of all profiles in the FOV. While it is mathematically the most consistent way to calculate the inverted physical atmosphere, it introduces additional degrees of freedom in the solution, because the sum of a large number of spectra, and not each individual spectrum is fitted to the data. This implies that a change in the spectrum in one pixel can be effectively ``hidden'' in the degraded data by an opposite change in the spectrum of the adjacent pixel. In addition, the current implementation is unable to deal with non-uniform instrumental properties, currently limiting its application to data from space-based instruments.

A selection of inverted quantities obtained using this method is shown in Figure~\ref{fig:map}, where 
a comparison is made between a coupled inversion and a traditional ``1D'' inversion, where in strumental effects are ignored. Despite the additional degrees of freedom, the coupled inversion maps look much more consistent, and show more detail than the 1D result. 

Further treatment of instrumental degradation will need to include the capability to deal with changing degradation conditions, to be able to invert ground based data. The framework for this already exists, but still requires impractical numerical resources to implement.

\subsection{Optimization}

At present, there are a number of basic methods in use for obtaining the most probable atmospheric structure, responsible for producing the observed spectra: classical downhill minimization, genetic algorithms and Bayesian methods.

\subsubsection{Genetic}

This method is loosely based on the principle of evolution by survival of the fittest, over a given number of generations, and is implemented using the PIKAIA algorithm \citepads{1995ApJS..101..309C} as the optimization method for the HELIX code. The optimization strategy consists of setting up a number of populations of trial solutions, each with an internal variation of the fit parameters. A given fraction of the populations is then eliminated, with a chance of survival that is mostly determined by the quality of the agreement with the observed data, usually based on a least squares metric. The remaining solutions are then duplicated, and random changes are introduced, thus giving rise to a new generation. 

There are a number of advantages, such as a guarantee that the global minimum will be found if enough generations are allowed to pass, and automatic detection of degeneracies in parameter space. It also only requires the forward calculation of the synthetic line profiles, and not the derivative of the merit function to the atmospheric fit parameters, which makes it very flexible. 

One major disadvantage is that the number of trials per generation that is necessary to probe all fit parameters independently is dependent on the number of atmospheric fit parameters, so that the numerical cost grows very rapidly with the complexity of the fitted atmosphere. This has limited the use of this method to relatively simple slab and Milne-Eddington atmospheres. 

\subsubsection{Bayesian/MCMC}

Bayesian inversions \citepads{2007A&A...476..959A} are a probabilistic approach to inversion, in the sense that they determine for a distribution of trial solutions the posterior, that is: a distribution of the probability that any of the trial solutions is in agreement with the observed data. The most probable value of the solution is the one with the largest probability, but the uncertainty in the parameter values can be extracted from the posterior as well.

The key ingredient for this type of inversion is a method that generates a set of trial solutions that efficiently samples the posterior. A brute force grid computation will obviously do the job, but scales poorly with the number of fit parameters. The Markov Chain Monte Carlo (MCMC, f.i. \citeads{metropolis_53}) method is one way to reduce this effort, and has been used successfully applied in recent years \citepads{2011ApJ...731...27A,2014A&A...572A..98A}. MCMC is not unlike the genetic method, with the difference that new generations are based on old ones in a different way, that most efficiently samples the posterior, for which an estimate of the posterior is used.

The main limitation of this method is that to sample the posterior, many forward calculations are needed, which limits the applicability of the method to relatively simple and fast atmospheric models.

\subsubsection{Database/PCA}

This method is the most direct, but therefore also the most thorough. The aim is to quantify the spectra in some way that can be used to categorize them. This can be done by using the intensity values in each wavelength bin of the data, but also by using the value of the inner product of the data with some basis of spectral basis functions. An optimal set of such basis functions can be produced using Principal Components (PCA), a small number of which is usually sufficient to describe each spectral profile to the noise limit  (\citeads{2000A&A...355..759R}; \citeads{2001ApJ...553..949S}). The objective is now to build an exhaustive database of spectra of all possible atmospheres, which are all described by the coefficients used to categorize the spectra. This database then can be used as a look-up table for an observed spectrum, for which the database lookup then returns all possible atmospheres that yield a sufficiently similar spectrum.

The result is thus a distribution of atmospheric values, not a specific value. If the distribution is normal, an error may be specified, as is often done for gradient search minimizations. However, frequently the distributions are far from normal, and there can be co-dependencies between parameters, that indicate there are degeneracies in the model.

Clearly, these methods return a lot of information that can be used to evaluate the result, but this comes at the cost of having to calculate a database that is complete. This limits the applicability of such methods to relatively simple atmospheres, since the dimensionality of the database quickly grows beyond the capabilities of available computational resources.

\subsubsection{Gradient search}

This is a classic method for finding the optimum solution to a given problem, for which a large body of mathematical work exists (see for instance \citeads{press1992}). The method is really a collection of methods, most of which are very efficient, in that they require only a few trial calculations of the synthetic spectra to converge, but they suffer from the problem that convergence to the global minimum can only be guaranteed if the merit function is monotonic. In addition, the gradient of the merit function to the atmospheric fit parameters is required, which can be difficult or costly to compute. In most inversion codes, the gradient is not explicitly calculated, but instead the response functions of the atmosphere are used to drive the solution in the direction of the minimum, by eliminating the remaining difference between the forward calculated spectrum and the observed one. It is relatively easy to show that when a $\ell_2$-norm is used, the gradient of the merit function depends on the response function. Assuming that the parameters of our model atmosphere are encoded in a vector $\mathbi{x} = {x_1, x_2, ..., x_i}$, for an observation $\mathbi{O}(\lambda)$ consisting of $N_\lambda$ wavelengths in the four Stokes parameters, the merit function ($\chi^2$) and the gradient relative to a parameter $x_i$  are given by
\begin{eqnarray}
\chi^2 &=& \frac{1}{4N_\lambda}\sum_{j<4N_\lambda} \frac{({S}_j(\mathbi{x}) - {O}_j)^2 }{\sigma_j^2}	\\
\frac{\partial \chi^2}{\partial x_i} &=&  \frac{1}{2N_\lambda}\sum_{j<4N_\lambda} \frac{({S}_j(\mathbi{x}) - {O}_j) }{\sigma_j^2} 
\underbrace{\frac{\partial {S}_j(\mathbi{x})}{\partial x_i}}_{\mathbi{R}(x_i)}
 \label{eq:grad}
\end{eqnarray}
where $\mathbi{S}(\mathbi{x})$ are the synthetic profiles and $\sigma$ is the noise of the observations. The last term in Eq.~(\ref{eq:grad}) is the response function of $\mathbi{S}$ to perturbations in $x_i$.

\begin{figure*}[hbt]
    \centering
    \includegraphics[width=0.8\textwidth]{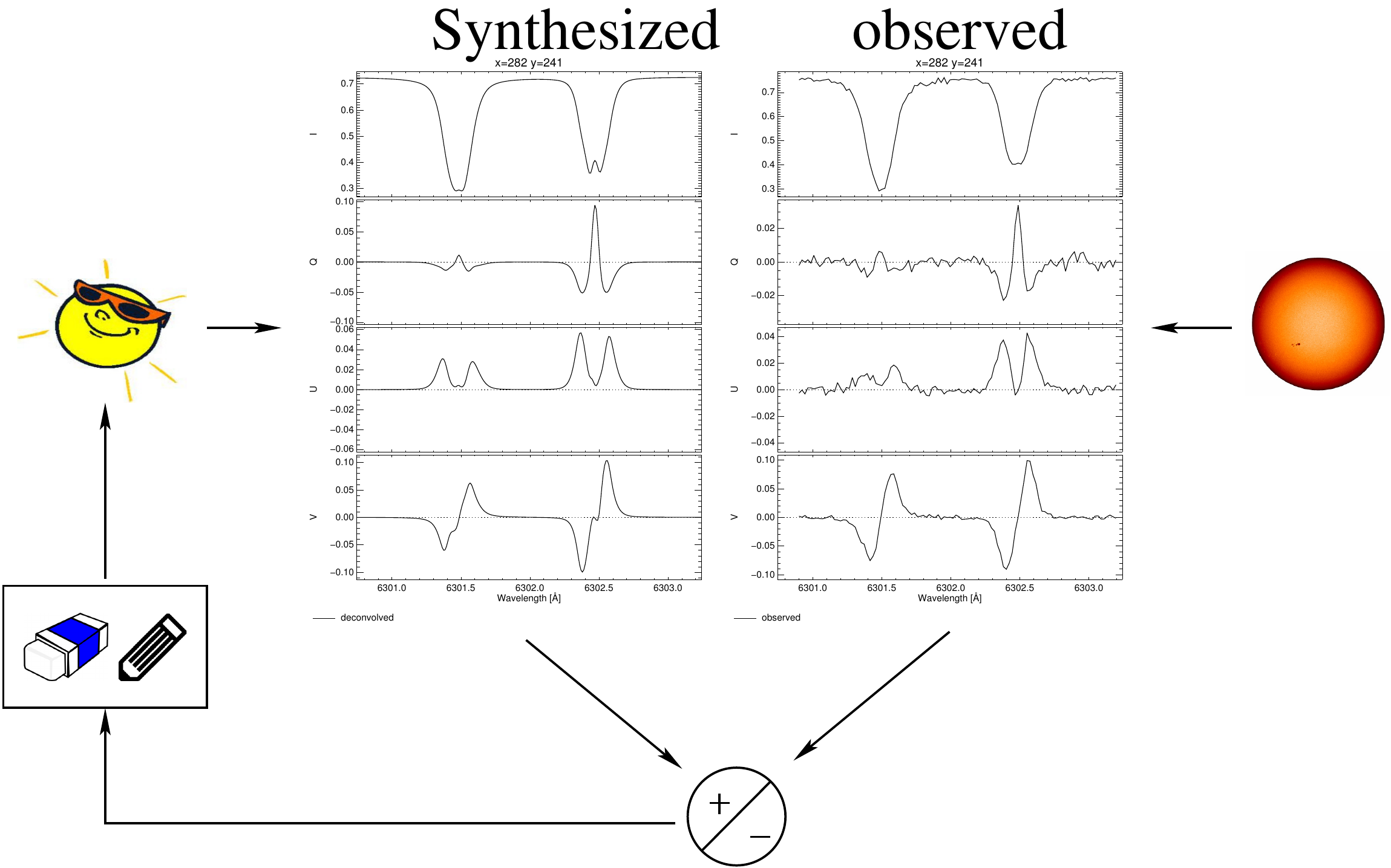}
    \includegraphics[width=0.8\textwidth]{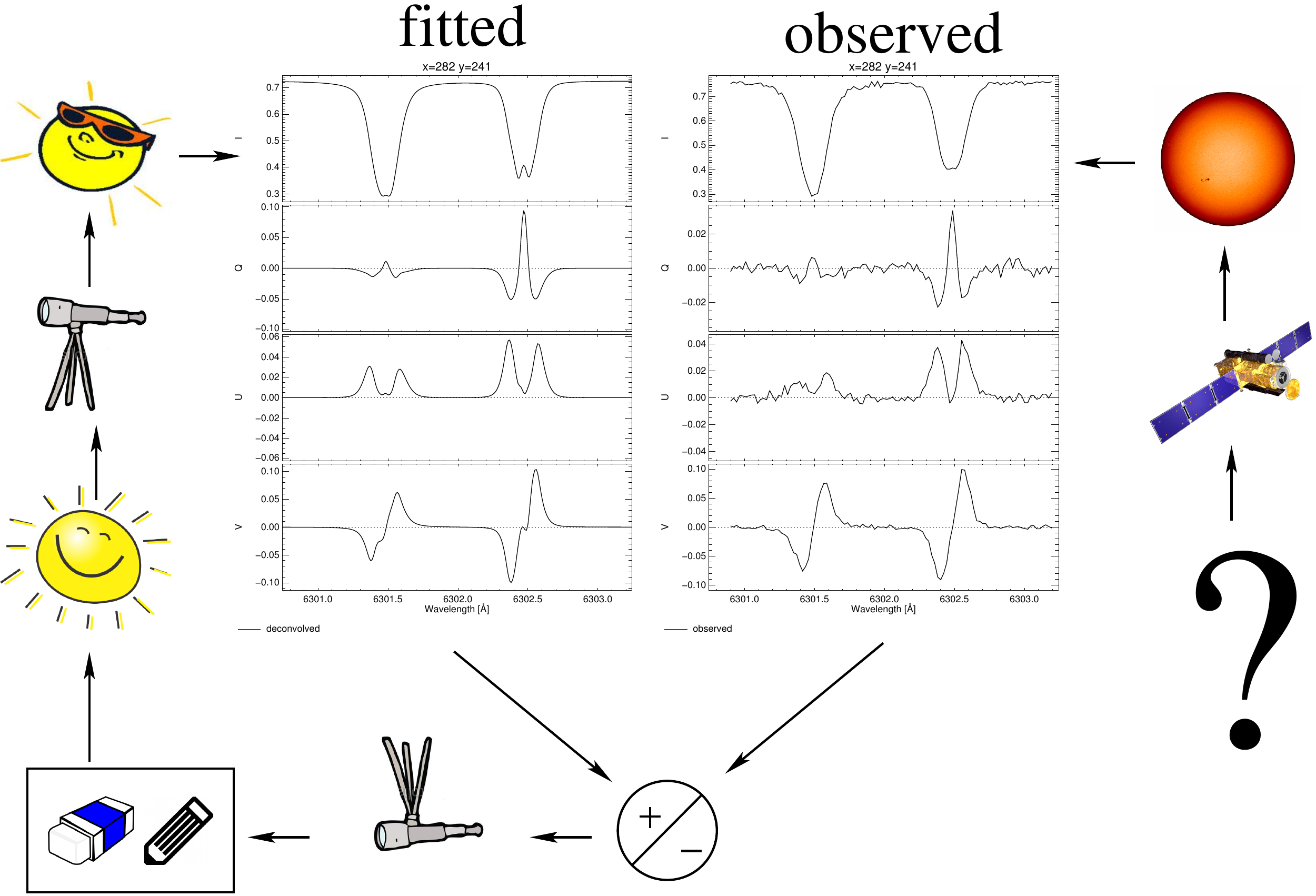}
  \caption{Gradient search optimization. {\bf Top}: single spectrum, without instrumental effects. {\bf Bottom}: coupled inversion, with instrumental effects considered.}
  \label{fig:invert}
\end{figure*}

The most common gradient search minimization algorithm in use is the Levenberg-Marquardt method \citepads{1944QoAM...2..164L,1963SoAM...11..431M}, which is a mix of the minimization of the linearized problem and a local gradient search that kicks in when the forward problem is very non-linear. It is at the heart of all commonly used height dependent atmospheric inversion codes, and has a robust convergence, as long as the derivatives of the synthetic spectrum with respect to the atmospheric fit parameters can be efficiently obtained. Figure~\ref{fig:invert} (left) shows the basic scheme, where the difference between the observed and synthetic spectrum are used to drive corrections to the current estimate of the atmosphere. 

It is possible to adapt the gradient search algorithm to search for the minimum of the difference between the spatially degraded synthetic profiles and the data, instead of using the synthetic profiles directly, by applying the effect 
that instruments have on the spectra also to the response functions, before the correction is calculated. Although this requires some additional effort in the computation of the correction, the gradient search method is currently the only method that has been successfully applied to simultaneously invert more than $10^6$ atmospheric variables \citepads{2012A&A...548A...5V}. Fig.~\ref{fig:invert} illustrates gradient search algorithm for spatially coupled inversions and traditional pixel-by-pixel inversions.

%% file: obs.tex
\section{Main diagnostics in the chromosphere}\label{sec:diagno}
Observing the chromosphere and the transition region poses a challenge for solar
physicists. There is a very limited selection of spectral lines that
have sufficient opacity to sample the chromosphere. 

Ground based
observations are limited to visible and infrared wavelengths, and
the most common chromospheric diagnostics are perhaps the
\ion{Ca}{ii}~H \& K lines ($\lambda 3934$ \& $\lambda 3968$), the
\ion{H}{i}~$6563$ line (H$\alpha$), the \ion{Ca}{ii} infrared triplet
lines (IR, $\lambda 8489$, $\lambda 8542$ \& $\lambda 8662$) and the
\ion{He}{i}~$5876$ (D$_3$) and $\lambda10830$ lines. 

The ultra-violet includes a list of interesting lines that must be
observed from space. These diagnostics are highly unexplored compared
with those in the visible and infrared: the \ion{H}{i}~$\lambda 1216$ line
(L$\alpha$), \ion{Mg}{ii}~h \& k lines ($\lambda 2796$ \& $\lambda
2804$), the \ion{He}{i}~$584$ line, the \ion{He}{ii}~$304$ line,
\ion{Si}{iv}~$\lambda 1394$ and $\lambda 1403$, and
many more. In table~\ref{tab:lines} we summarize modelling requirements for some chromospheric diagnostics.
\begin{figure*}
\centering
\includegraphics[width=\textwidth, trim= 1.2cm 0 0.7cm 0, clip]{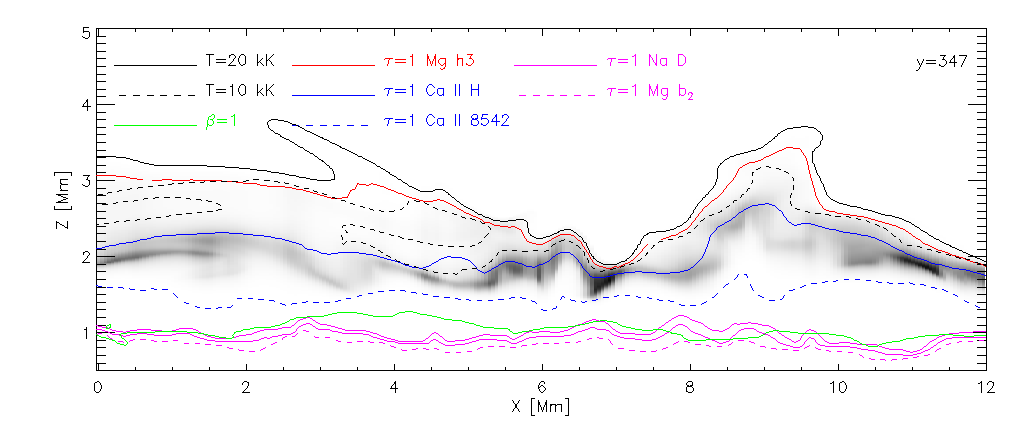}
\caption{Slice from a 3D MHD simulation indicating the $\tau=1$ surface at line center for \ion{Mg}{ii}~h$_3$~\&~b$_2$, \ion{Ca}{ii}~H \& $\lambda 8542$, and \ion{Na}{i}~D. The contribution function of the \ion{He}{i}~$\lambda 10830$ line is rendered in grey-scale because it is optically thin. Courtesy of M. Carlsson (Institute for theoretical
astrophysics, University of Oslo).}
\end{figure*}

The \ion{Ca}{ii}~H \& K lines are particularly sensitive to
temperature because the Planck function is very steep at those
wavelengths. Thus, their broad wings can act as remarkable thermometers for the photosphere and the lower chromosphere. These lines have slightly higher Land\'e factor than the IR
triplet lines ($g_{3933}=1.16$, $g_{3968} = 1.33$) but the lower photon count in the UV and the dependency of the Zeeman splitting with $\lambda^2$ makes them effectively less sensitive to magnetic fields (see Eq.~\ref{eq:vb}). 
Partial redistribution effects (PRD) must be
included to reproduce the intensity profile close to
line center (\citeads{1989A&A...216..310U}). The real advantage of
these lines is that they can be observed from the ground at the
highest resolution that a telescope can deliver as the diffraction
limit of a telescope scales linearly with $\lambda$. The importance of
3D non-LTE effects remains to be assessed in a similar way as
\citetads{2012A&A...543A..34D} did for the $\lambda 8542$ line. 

The \ion{Mg}{ii}~h \& k lines sample the transition region, providing
valuable information of the physical coupling between the chromosphere
and the corona (\citeads{1997SoPh..172..109U}). With the launch in 2013 of NASA's Interface Region
Imaging Spectrograph (IRIS, \citeads{2014SoPh..289.2733D}), these
lines have become particularly interesting for the solar
community (\citeads{2013ApJ...772...90L};
\citeads{2013ApJ...778..143P}), although without measuring
polarization. However, these lines could be
combined with the \ion{Ca}{ii} IR triplet lines to constrain the
temperature stratification of the solar chromosphere and to measure
the magnetic field in the chromosphere. PRD of scattered photons must
be used to model these lines.

The H$\alpha$ line has been considered for a very long time one of the
best chromospheric diagnostics (see comments by
\citeads{2007ASPC..368...27R}). 
It is well suited for the measurement
of chromospheric velocities because the line profile is observed in
absorption in most chromospheric conditions (not always
though). \citetads{2009A&A...503..577C} showed that the width of the
Gaussian core of the H$\alpha$ line is strongly affected by the
chromospheric temperature because the hydrogen atom is very
light. However, this line is not well suited for
current implementations of non-LTE inversions because to reproduce the observed
intensities, the radiation field must be evaluated in 3D dimensions
(\citeads{2012ApJ...749..136L}). H$\alpha$ is not an ideal diagnostic
for chromospheric magnetic fields because its response to the Zeeman
effect is originated in the photosphere
(\citeads{2004ApJ...603L.129S}), although the line is sensitive to the
Hanle effect close to the line center (e.g., \citeads{2010ApJ...711L.133S}).
 
The \ion{Ca}{ii}~IR triplet
lines have a relatively high sensitivity to magnetic fields
(with Land\'e factors $g_{8498}=1.07$, $g_{8542}=1.10$, $g_{8662}=0.87$), and they
can be modelled assuming CRD (see \citeads{1989A&A...216..310U}) and
statistical-equilibrium (\citeads{2011A&A...528A...1W}). Furthermore,
these lines are very sensitive to the temperature stratification of
the atmosphere and encode information about velocity
gradients. Therefore,
most, if not all, depth-stratified non-LTE inversions in the
chromosphere have been carried out using the \ion{Ca}{ii}~IR triplet
lines in active regions (). In the quiet-Sun, scattering polarization and
the Hanle effect must be included to describe the linear polarization signals
(\citeads{2010ApJ...722.1416M};
\citeads{2013ApJ...764...40C}). Recently,
\citetads{2014ApJ...784L..17L} showed that the effect of isotopic
splitting must be included in the inversions to properly retrieve the
velocity stratification.
\begin{figure*}
\centering
\includegraphics[width=0.9\textwidth]{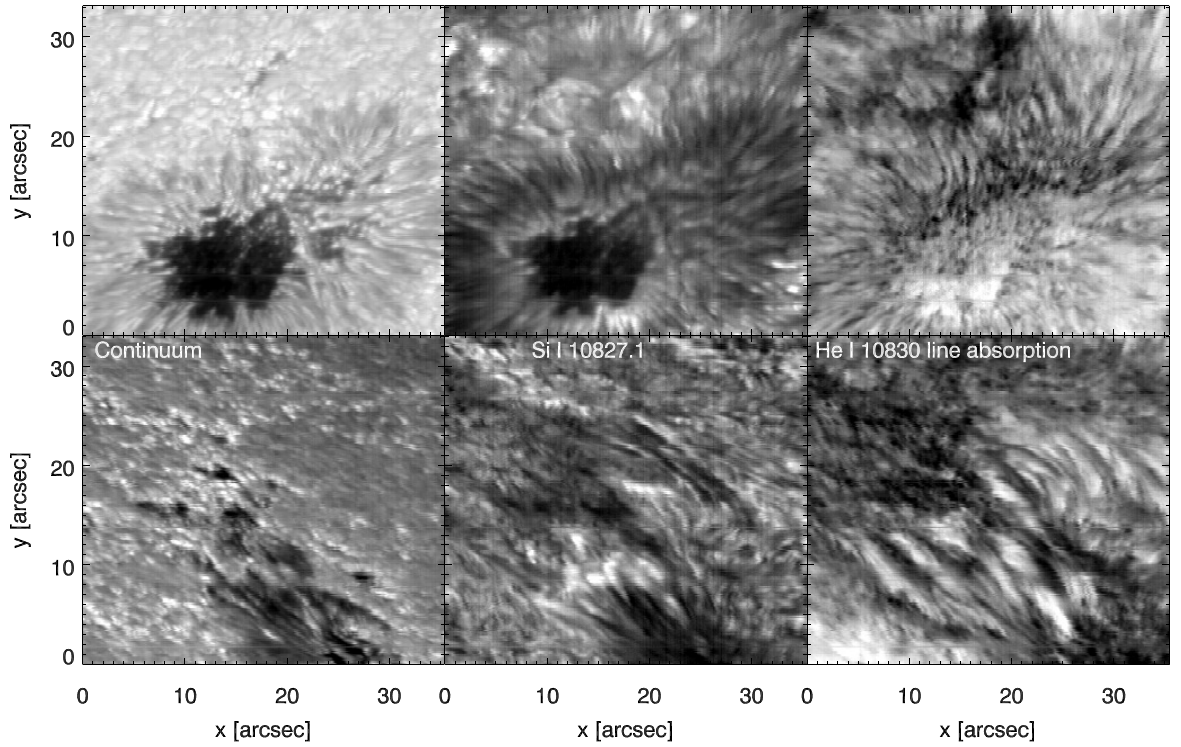}
\caption{Raster-scan observation of AR12394 (top) and AR12390 (bottom) on 2014-08-01 at 1083~nm with the TRIPPEL spectrograph at the Swedish 1-m Solar Telescope. From left to right the panels show images in the continuum, in \ion{Si}{i}~$\lambda 10827$ line center and in \ion{He}{i}~$\lambda 10830$ line absortion. Courtesy of J. Joshi, T. Libbrecht, G. Sliepen and D. Kiselman (Institute for Solar Physics, Stockholm University).}
\label{fig:trippel}
\end{figure*}

The physical processes
involved in the formation of the \ion{He}{i}~D$_3$ and $\lambda10830$~lines are significantly different than those
of the H$\alpha$ line and of the aforementioned \ion{Ca}{ii}
lines (see \citeads{1939ApJ....89..673G}; \citeads{1997ApJ...489..375A};
\citeads{2007ApJ...655..642T}; \citeads{2008ApJ...677..742C}; \citeads{2014ApJ...784...30G}; \citeads{2016arXiv160800838L}). The \ion{He}{i} atom consists of two spectral series of
singlets (para-helium) and triplets (ortho-helium). The ground state
of the atom is a singlet level, but the D$_3$ and the $\lambda10830$ transitions
take place between two levels of the triplet system. Since
transitions between the para and the ortho levels are forbidden by the
dipole selection rules and collisional rates in the chromosphere are
very low, the only way to populate the ortho-levels is through
photo-ionization from the para-helium system followed by a
recombination into the ortho-system. The electrons cascade down
through the levels of the ortho-system until the ground meta-stable
level of the ortho-helium system. The ionization potential of the
para-system is 24.6~eV, which means that the photoionization of the
\ion{He}{i} atom can only be effectively done with UV photons
($\lambda < 504$~\AA). Such radiation can only originate in the solar
corona, which makes these lines particularly interesting because they are
sensitive to both chromospheric and coronal physical conditions; see observations with the TRIPPEL spectrograph (\citeads{2011A&A...535A..14K}) at the Swedish 1-m Solar Telescope (SST, \citeads{2003SPIE.4853..341S}) in Fig.~\ref{fig:trippel} and with Gregor/GRIS (\citeads{2012ASPC..463..365S}; \citeads{2007ASPC..368..611C}) in \citetads{2015SSRv..tmp..115L}. Since
UV photons can only penetrate a thin layer of the chromosphere before
they are absorbed, these lines are assumed to form in a relatively
thin and localized range of heights, making the constant slab or
Milne-Eddington descriptions applicable.

\begin{table*}[!ht]
\centering
\begin{tabular}{|l|c|c|c|c|}
\hline
{\bf Line(s)} & \textbf{Scattered photons} & \textbf{Zeeman/Hanle} & \textbf{Geometry} & Ionization \\
\hline
\ion{Ca}{ii}~H \& K & PRD & Zeeman (AR), Hanle (QS) & 1.5D & stat. equilibrium\\
H$\alpha$ & CRD & Zeeman (AR), Hanle (QS) & 3D & non-equilibrium \\
\ion{Ca}{ii} IR triplet & CRD & Zeeman (AR), Hanle (QS) & 1.5D & stat. equilibrium \\
\ion{Mg}{ii} h \& k & PRD & Zeeman, Hanle (k line) & 1.5D & stat. equilibrium \\
\ion{He}{i} D$_3$ \& $\lambda 10830$ & CRD (?) & Zeeman + Hanle & 3D (?) & non-equilibrium \\
\hline
\end{tabular}\label{tab:lines}

\end{table*}

\section{Inversions in the chromosphere: a selection of results}\label{sec:chrodiag}
This volume includes dedicated chapters about sunspots (Rempel \& Scharmer, same issue) and about photospheric magnetic fields in the quiet-Sun (Schussler et al., same issue). Therefore, we have chosen to focus on a small selection of results obtained from chromospheric studies, and more specifically on those concerning inference of magnetic fields. 

 The first attempts
to \emph{invert} chromospheric data can be tracked back to the VAL and
FAL atmospheres (see \citeads{1981ApJS...45..635V} and
\citeads{1993ApJ...406..319F}). Those studies used spatially-averaged
spectra, including many chromospheric lines from different species
(atoms and ions), to derive models that
could reproduce the observations. Nowadays, the VAL and FAL atmospheres are still used to derive
radiative fluxes in the chromosphere or as reference chromospheric
models. However, the chromosphere is highly dynamic and finely
structured. These 1D models cannot catch those two inherent properties
of the chromosphere, and it is now widely accepted that
spatio-temporal models must be used to characterize this region of the Sun.

\begin{figure*}
\centering
\includegraphics[width=0.8\textwidth,trim = 0 0 0.4cm 0, clip]{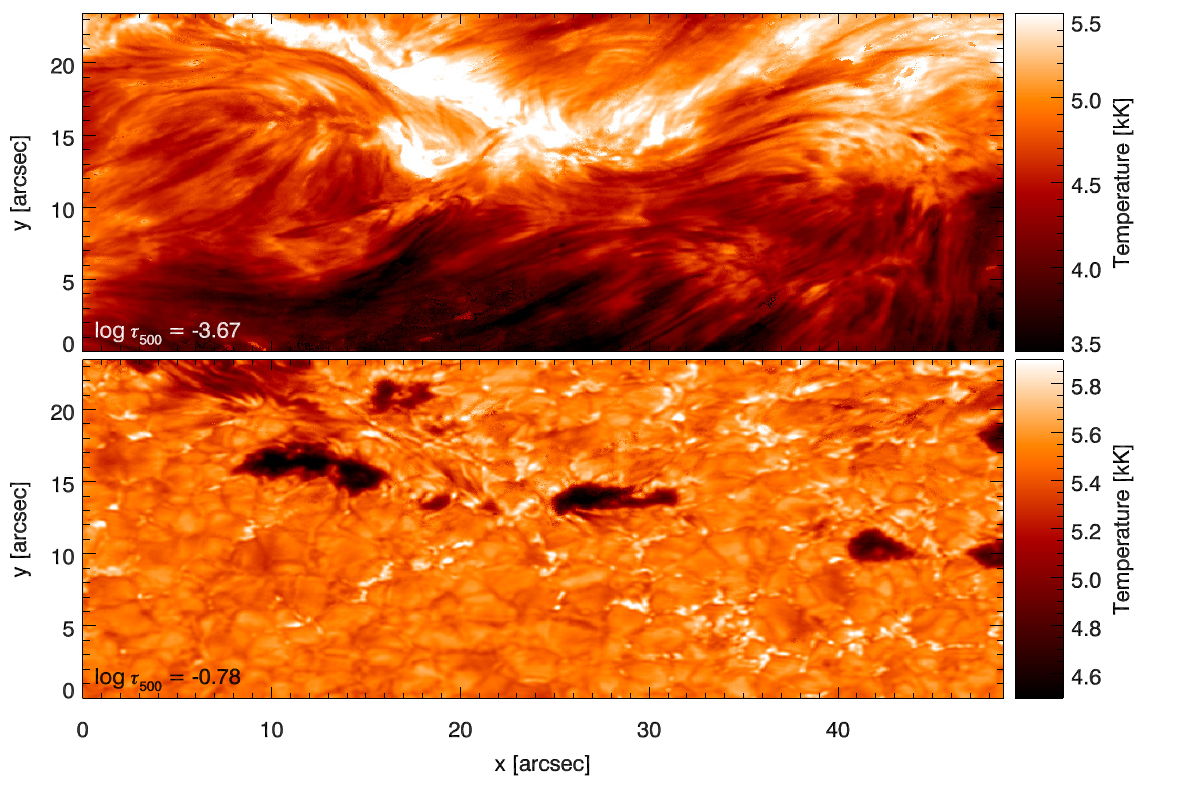}
\caption{Inferred chromospheric (top) and photospheric (bottom) temperatures from a non-LTE inversion using observations in the \ion{Ca}{ii}~$\lambda8542$ line acquired with CRISP (\citeads{2006A&A...447.1111S}; \citeads{2008ApJ...689L..69S}) at the SST. The temperature maps are given at fixed optical depths of $\log \tau_{500} = -3.67$ (top) and $\log \tau_{500} = -0.78$ (bottom).
A small patch from this inversions was used by \citetads{2015ApJ...810..145D} to study small scale magnetic flux emergence.}\label{fig:temp}
\end{figure*}

The following techniques have been particularly successful to derive
magnetic fields in the chromosphere:
\begin{enumerate}
        \item Depth-stratified non-LTE inversions, assuming
        statistical equilibrium and complete redistribution in angle
        and frequency of scattered photons (CRD). \citetads{2000ApJ...530..977S} developed inversion methods that
are used nowadays (\citeads{2015A&A...577A...7S}) to derive spatio-temporal model atmospheres
including a photosphere and a chromosphere (see detailed temperature maps in Fig.~\ref{fig:temp}). 
\label{it:nlte}
        \item Milne-Eddington atmosphere or constant physical slab
        inversions. These inversions work directly with the parameters
        of the radiative transfer equation, and therefore only provide
        \emph{direct} information about the line-of-sight velocity and
        magnetic field vector.\label{it:me}
        \item The weak-field approximation can be used in most
        chromospheric situations to infer the magnetic field
        vector. Most chromospheric lines have a relatively low Land\'e
        factor and a relatively large Doppler width, making them
        suitable candidates for this simple approximation.
\end{enumerate}

\subsection{Depth-stratified non-LTE inversions} \label{sec:dep}
We focus now in approach \ref{it:nlte}. The statistical-equilibrium equations (Eq.~\ref{eq:stateq}) have an explicit dependency on the mean intensity
$\bar{J}$, which in current implementations is computed for each pixel assuming plane-parallel
geometry. In reality, if radiation is not coupled to the local
conditions of the plasma, horizontal radiative transfer can become
important, and the evaluation of the radiation field must be performed
in 3D. Therefore, this method is only suited for lines that can be
modelled assuming plane-parallel geometry. 

\noindent A number of  studies including  the inversion of \ion{Ca}{ii}~IR data can be found in the literature:

\emph{Sunspots:} Non-LTE inversions have allowed to identify the origin of umbral
flashes (UF) in sunspots and to derive
the chromospheric temperature increase during the flash phase
(\citeads{2000Sci...288.1396S}). The
polarity reversals observed in Stokes~$V$ during UF are induced by
emission in the intensity profile, not to actual changes in the
polarity of the magnetic field (\citeads{2013A&A...556A.115D}). In
fact, most of the variability that
is observed in monochromatic Stokes~$V$ images is produced by the
thermal properties of the plasma that leave an imprint in
Stokes~$V$, like Doppler motions and Doppler broadening. Similar
conclusions were drawn from the study of \ion{Ca}{ii}~H \& K data in
plage (\citeads{1990ApJ...361L..81M}). {The 3D structure of the magnetic field vector
has been retrieved by \citetads{2005ApJ...631L.167S}, who found a complex field configuration
with opposite-sign torsion that could be interpreted as opposite helicity flux ropes inside the same spot. In a second study
\citetads{2005ApJ...633L..57S} analyzed chromospheric heating induced by currents dissipation in the upper layers of the same sunspot.} One aspect that remains to be
investigated is the full depth stratification of the penumbra of
sunspots, and how the Evershed effect and the moat flow relate to the
inverse Evershed flow and the superpenumbra observed in the chromosphere.

\begin{figure*}
  \centering
  
\includegraphics[width=0.9\textwidth]{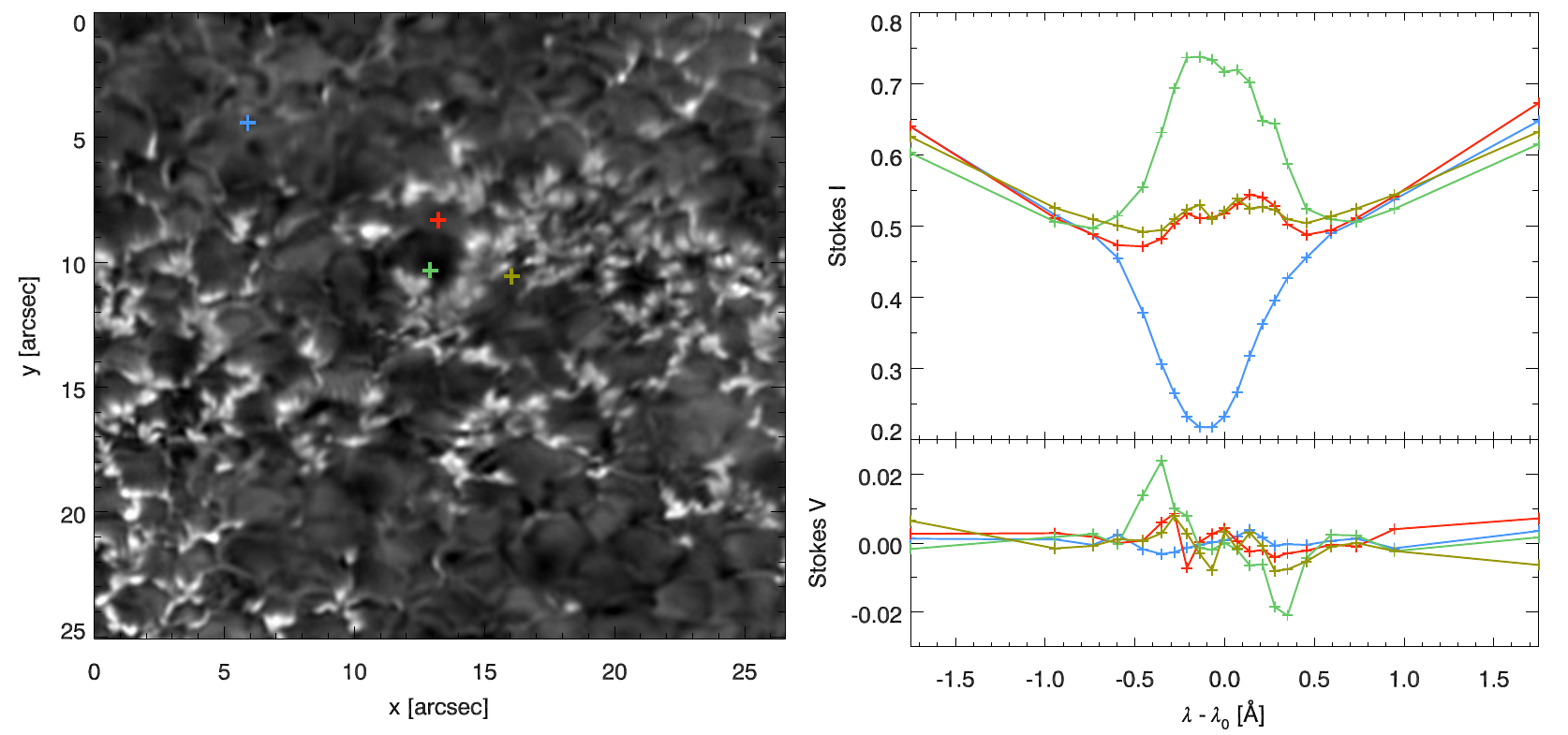}
\caption{Observation of plage in the \ion{Ca}{ii}~$\lambda 8542$ line
  recorded with the CRISP instrument at Swedish 1m Solar
  Telescope. \emph{Left:} Monochromatic image acquired in the wing of
  the line, showing bright-points embedded in photospheric
  granulation. The color markers indicate the locations of the
  spectra plotted on the \emph{right} panel, with the same color
  coding.}\label{fig:fprof}
\end{figure*}

\emph{Plage:} \citetads{1974SoPh...39...49S} discussed ad-hoc 1D models that could
reproduce some observed properties in plage profiles. However, the
first and probably only attempt to apply
field-dependent non-LTE inversions to a patch of network was carried out by
\citetads{2007ApJ...670..885P}, using slit-spectrograph observations
in the \ion{Ca}{ii}~$\lambda 8498$ and $\lambda 8542$ lines, recorded with
SPINOR (\citeads{2006SoPh..235...55S}). One of their main results is
the appearance of a field that expands with height, and becomes more
horizontal in the chromosphere, suggesting a canopy effect
(see flux-tube models by \citeads{1991A&A...250..220S},
or measurements of increasing magnetic field gradients in plage by
\citeads{1989A&A...222..311S}). However, their conclusions are based
in the visual appearance of the inverted field topology because they
did not have enough signal in Stokes $Q$ and $U$ to derive the
inclination of the magnetic field. This canopy effect has also been found in
3D MHD simulations (\citeads{2013ApJ...764L..11D}), which has a clear
diagnostic associated: the core of the \ion{Ca}{ii}~IR lines is
observed in emission or with greatly enhanced intensity (see
Fig.~\ref{fig:fprof}). It remains unclear what process heats the
chromosphere in this canopy, above the \emph{quiet} photosphere below, but
it must have a magnetic origin.

\emph{Quiet-Sun:} In quiet-Sun observations, the amplitudes of
Zeeman induced polarization (in most chromospheric lines) are similar
to those originated from
scattering polarization and from the Hanle effect. These amplitudes are
expected to be very low, of the order of $10^{-4}$ relative to the
continuum intensity. Therefore, the challenge to measure chromospheric
magnetic fields is twofold: first it is
very hard (probably impossible at the moment) to reach the required sensitivity
with high/medium spatial resolution, but even if such observations were
available, modelling the observed poarization profiles would be extremely hard. One of
the best estimates of the field strength in the chromosphere of quiet-Sun 
is given by \citetads{2004Natur.430..326T}.


\subsection{Constant slab inversions and Milne-Eddington
inversions}\label{sec:slab}

The first realistic
inversions of the D$_3$ line,  including a full treatment of
the Hanle effect and scattering polarization, were originally introduced by
\citetads{2002ApJ...575..529L}. To overcome the computational
challenge (at that time) of performing a full least-squares-fit for
this non-linear problem, they used a method based on PCA decomposition
and precomputed synthetic profiles
(\citeads{2000A&A...355..759R}). \citetads{2003Natur.425..692S}
studied the $\lambda 10830$ line in an
active region, including only Zeeman induced polarization in their
inversions. HAZEL was the first code that included
full spectral synthesis with Zeeman induced polarization, scattering
polarization and the Hanle effect (\citeads{2008ApJ...683..542A}), and Helix+
is now also capable of similar calculations (\citeads{2009ASPC..415..327L}).

The D$_3$ and $\lambda 10830$ lines have been used to study
prominences and filaments, spicules, fibrils, and flares. Here we
summarize some important results from the literature:

\emph{Fibrils:} To our knowledge, most of the studies carried out in
fibrils (using \ion{He}{i} lines) have focussed on measuring the direction of
the magnetic field and on assessing whether the magnetic field is aligned along
the direction defined by chromospheric fibrils. In particular, the fibrils observed
in the superpenumbra of sunspots seem to be aligned with the measured
direction of the magnetic field (\citeads{2013ApJ...768..111S}), although
\citetads{2011A&A...527L...8D} found a number of cases with
significant miss-alignment using the $\lambda 8542$ line. Further
studies of fibrils in the H$\alpha$ line using a 3D MHD simulation,
showed that fibrils are usually aligned with the magnetic field in the
horizontal direction, but that is not always the case vertically
(\citeads{2015ApJ...802..136L}).

\emph{Prominences:} Theoretical models of solar prominences predict
that the magnetic field must be highly inclined to support the
material inside the prominence. Observationally, most measurements of the magnetic field topology
in prominences have come from observations in \ion{He}{i}
lines ($\lambda5876$ and $\lambda 10830$). A difficulty that arrises from these observations is the
presence of ambiguities that are compatible with several orientations
of the magnetic field. The validity of a more inclined versus
a more vertical topology of the magnetic field has been greatly debated by the community (see latest results by \citeads{2005ApJ...622.1265C}; \citeads{2014A&A...566A..46O};
\citeads{2014A&A...569A..85S}; \citeads{2015ApJ...802....3M}). 

\begin{figure*}[hbt]
\centering
\includegraphics[width=0.8\textwidth, trim = 0.7cm 0.2cm 0.7cm 0.3cm, clip]{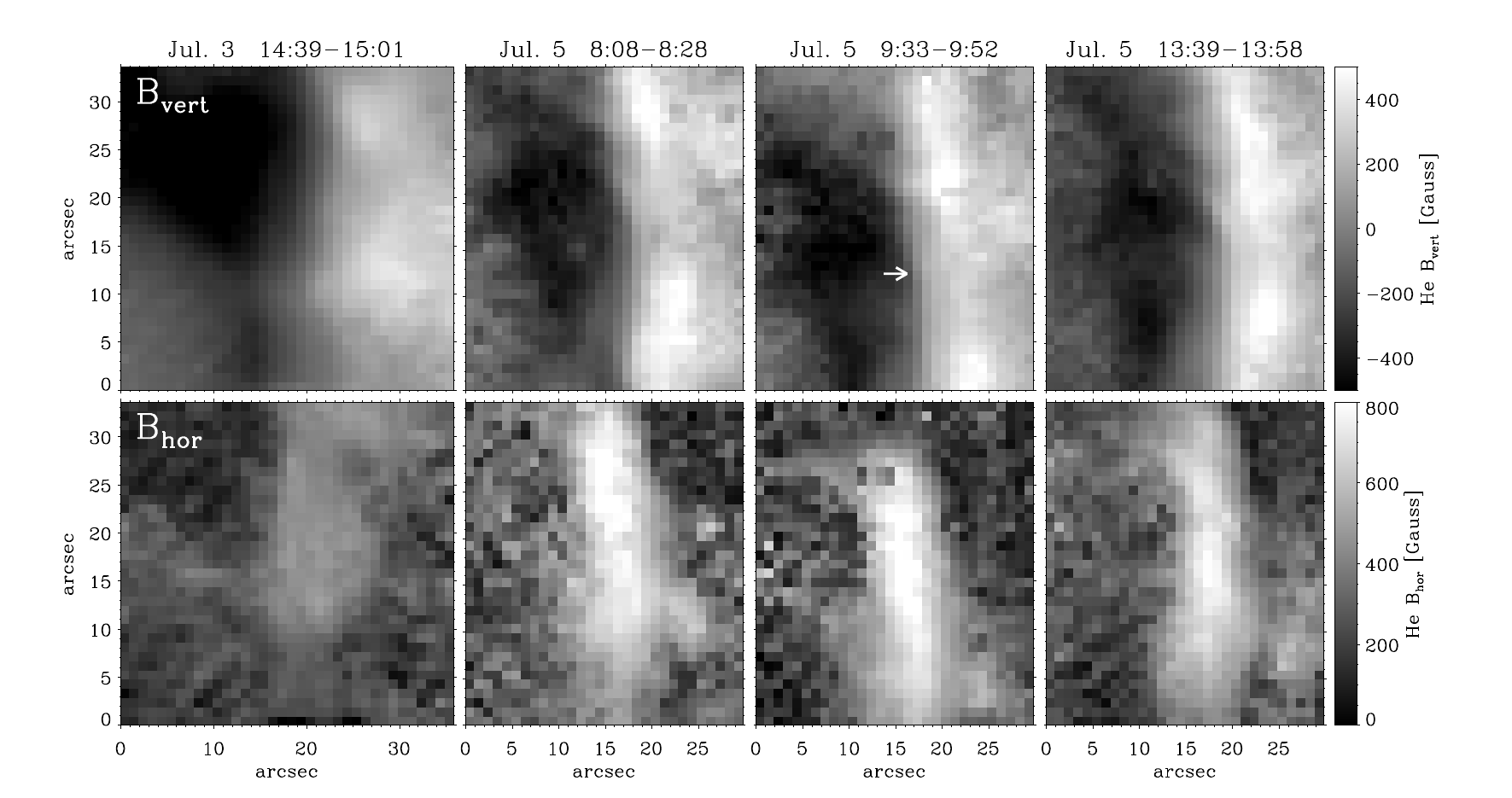}
\includegraphics[width=0.8\textwidth, trim = 0.7cm 0.2cm 0.7cm 0.3cm, clip]{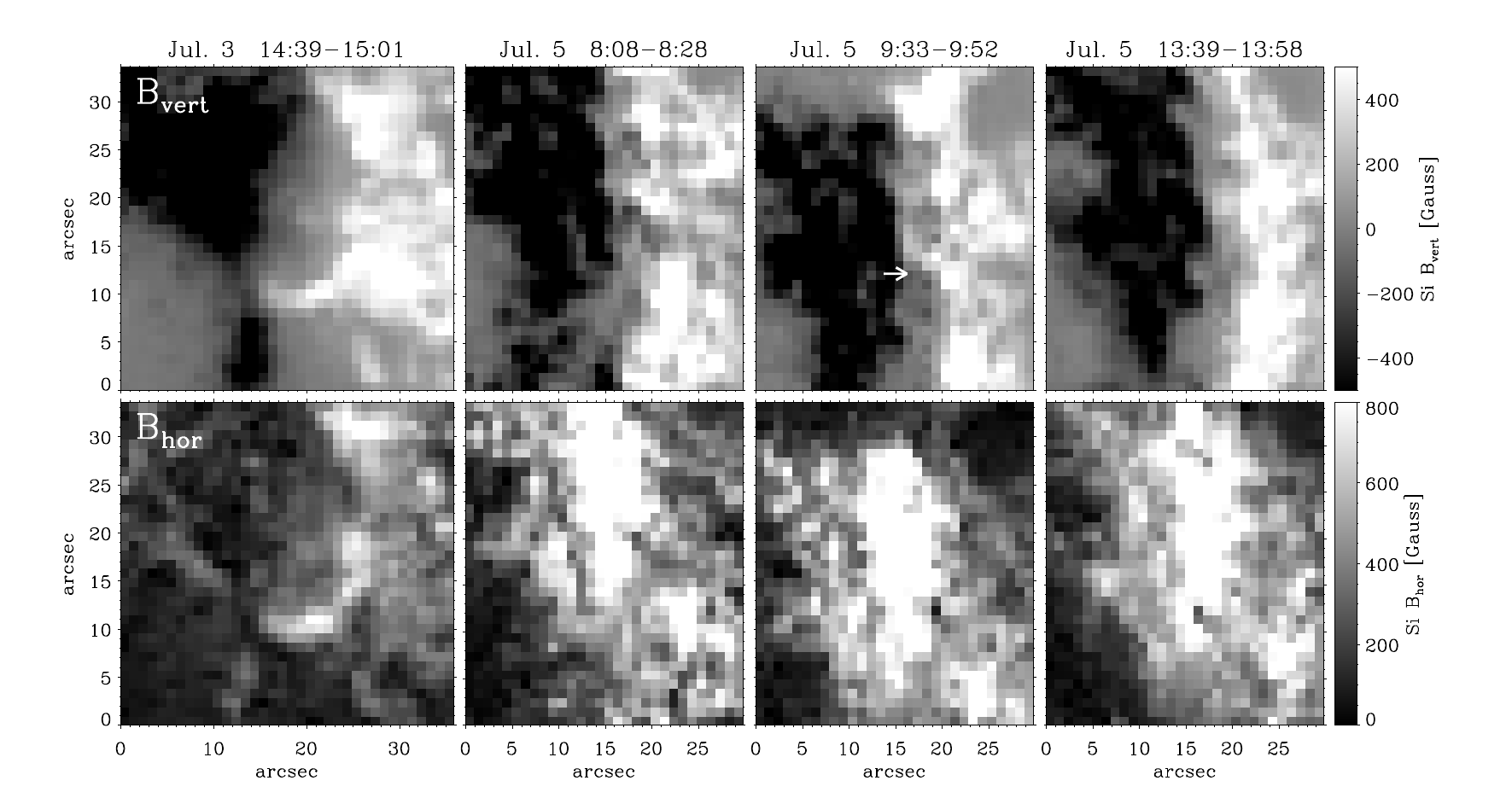}
\caption{Magnetic field in a filament as a function of time. The upper two rows correspond to chromospheric inversions of the \ion{He}{i}~$10830$ line, wheres the lower two rows correspond to photospheric inversions of the \ion{Si}{i}~$\lambda 10827$. For each set of inversions, the upper row corresponds to the vertical component, and the lower panel corresponds to the horizontal component. Reproduced from \citetads{2012A&A...539A.131K}, Fig.~11 and Fig.~12 combined.}\label{fig:fil}
\end{figure*}

\emph{Filaments:} Magnetic fields in filaments have been recently studied in the photosphere and in the chromosphere separately (\citeads{2005ApJ...622.1275L}; \citeads{2006A&A...456..725L}; \citeads{2006ApJ...642..554M}). For the first time, \citetads{2012A&A...539A.131K} carried out a detailed analysis of simultaneous photospheric and chromospheric spectropolarimetric observations, and they were capable of inferring the magnetic field vector in the photosphere (using the \ion{Si}{i}~$\lambda 10827$ line and in the chromosphere (using the \ion{He}{i}~$\lambda 10830$ line). These studies have shown that the field in the chromosphere has a strong horizontal component along the spine of the filament (up to 800~G), whereas the vertical component of the field has typical values of 400-500~G and are placed on both sides of the spine (see Fig.~\ref{fig:fil}). 

\emph{Spicules:} Polarimetric observations of spicules outside the
limb have been used to derive the magnetic field orientation
(\citeads{2005A&A...436..325L}; \citeads{2010ApJ...708.1579C}),
including the variation of the magnetic field with height
(\citeads{2015ApJ...803L..18O}). However, to achieve a high enough signal to
noise ratio, these observations have usually been integrated for
periods of 10 to 45 minutes. Unfortunately, the chromosphere is highly
dynamic and those measurements reflect the average magnetic field vector
over that integration period. 

\subsection{The weak-field approximation in the chromosphere}\label{sec:wfa}
The weak field approximation (WFA hereafter) poses a very fast method to estimate the
magnetic field vector from full-Stokes observations of magnetically
sensitive lines (\citeads{1967AnAp...30..257R}). It is applicable when
the Doppler width of the line ($\Delta\lambda_D$) is much larger than
the Zeeman splitting of the line ($\Delta \lambda_B$) {and if the magnetic field strength, 
field inclination and line-of-sight velocity are close to constant as a function of depth in the atmosphere}. Under those
conditions, Stokes~$Q$, $U$ and $V$ can be written as a function of
constant magnetic field vector components (with depth) and the first and second derivatives of
the intensity profile (see summary in \citeads{2004ASSL..307.....L}).

Most chromospheric lines have a relatively low Land\'e factor and are
usually broader than photospheric lines.
Therefore, it is relatively easy to satisfy the conditions mentioned
above, except that many chromospheric spectral lines are sensitive to
a large range of regimes, and can sample photospheric and
chromospheric conditions
simultaneously. However, this inconvenience can be easily overcome by
isolating a small wavelength range where the line can be assumed to
sample one atmospheric regime. We note that \citetads{2009ApJ...706..148W}
proposed a hybrid \emph{center of gravity} method to derive depth
information of the magnetic field. \citet{jennerholm} used synthetic observations in the $\lambda 8542$ line, computed from a 3D MHD simulation to assess the usability of the WFA. Their results show good agreement between the inferred components of the magnetic field and that in the simulation.

The weak field approximation has been used to study the evolution of magnetic fields in
flares (\citeads{2012SoPh..280...69H}), in a sunspot umbra
(\citeads{2013A&A...556A.115D}) and in plage
(\citeads{2007ApJ...663.1386P}), from observations in the
\ion{Ca}{ii}~$8542$~\AA\ line.

\begin{figure*}
\centering
\includegraphics[width=0.75\textwidth, trim = 0.2cm 0 0.3cm 0, clip]{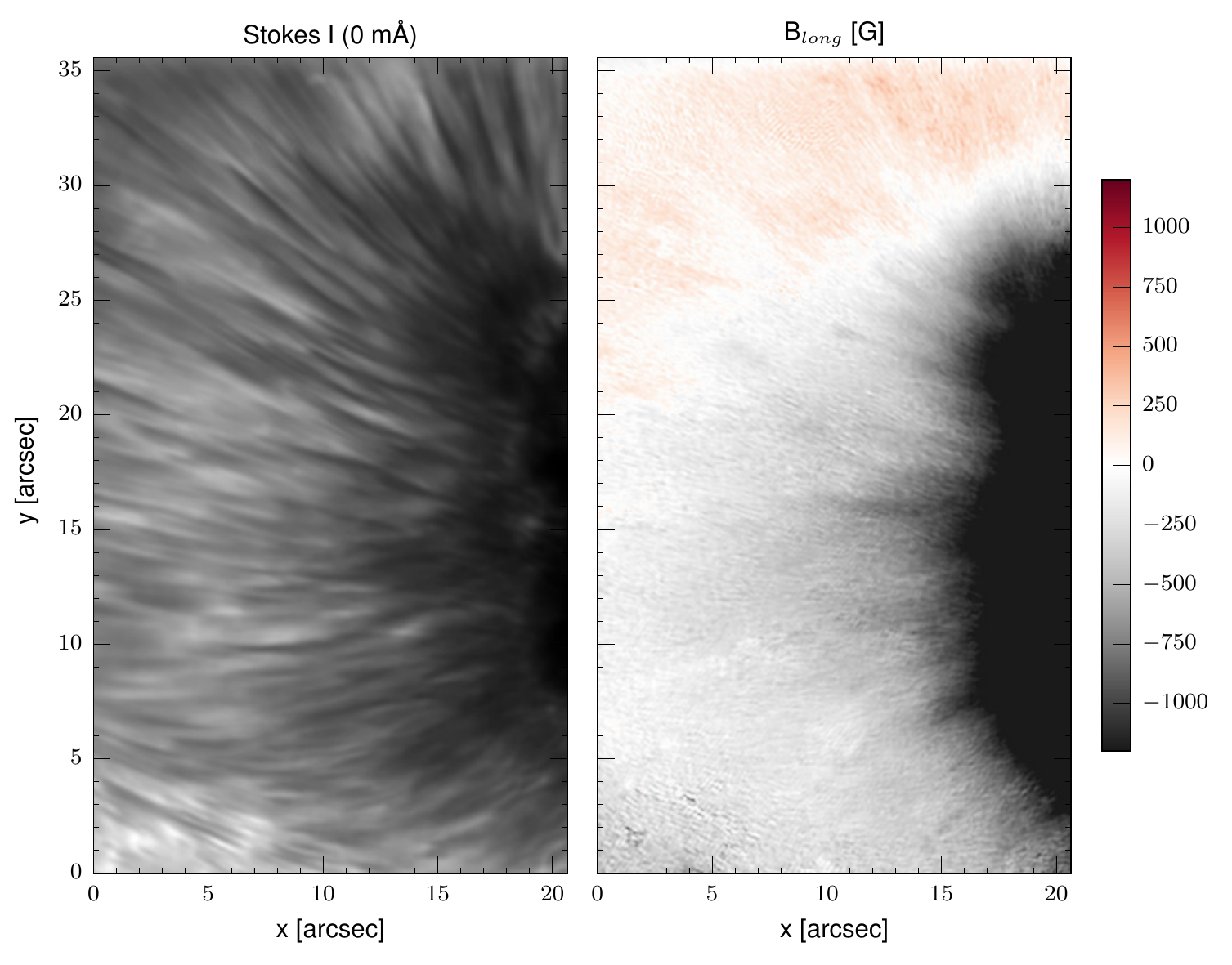}
\caption{SST/CRISP spectropolarimetric observation of AR1793 on
  2013-07-22 (08:51:55UT) in the
  \ion{Ca}{ii}~$8542$~\AA\ line. \emph{Left:} umbra and penumbra of a
  sunspot (AR1793) at the line center. \emph{Right:} the corresponding line of sight component of
the magnetic field vector, computed with the WFA.}\label{fig:wfa}
\end{figure*}

%% file: future.tex
\subsection{Structuring of magnetic fields in the chromosphere}
In \S\ref{sec:dep} and \S\ref{sec:slab} we have summarized a number of studies that have inferred magnetic fields, in most cases targeting particular features or phenomena in the chromosphere. The question of understanding how is the magnetic field topology in the chromosphere seems to persist. Recent developments in solar instrumentation have allowed to acquire spectropolarimetric observations with very high-spatial resolution (see summary by Kleint \& Gandofer, this issue), and monochromatic images in all the Stokes parameters can appear very finely structured. Particularly in the chromosphere, this fine structures may be (wrongly) assumed to indicate a similar topology of the magnetic fields, although magnetic fields may govern the intricate force balance in the chromosphere. This fine structure is likely to contain strong \emph{imprints} of Doppler motions, line broadening and opacity variations and corrugations, and it should not be assumed to resemble the magnetic field topology (similarly discussed by \citeads{2006ASPC..354..259J}). In the chromosphere gas pressure is several orders of magnitude lower than in the photosphere, whereas magnetic fields are usually one order of magnitude lower. While the photosphere is dominated by gas pressure gradients, that dictate the structuring of magnetic fields, in the chromosphere this is not clearly the case.

Fig.~\ref{fig:wfa} illustrates a highly structured Stokes~$I$ image close to line center in the $\lambda 8542$ line (left panel), but the inferred l.o.s. component of the magnetic field is much smoother, and similar results are obtained, e.g., in filaments (see Fig. 11 in \citeads{2012A&A...539A.131K}). Similarly, \citetads{2013A&A...556A.115D} found that most of the variability observed as a function of time in umbral flashes can be explained with an almost constant magnetic field (at each location). LTE inversions in the photosphere in sunspots and plage regions also reveal that the magnetic field rapidly expands as a function of height and becomes relatively smooth (e.g., \citeads{2013A&A...553A..63S}; \citeads{2015A&A...576A..27B}), supporting this scenario. 

 \emph{Therefore, magnetic fields in the chromosphere must be smoother than the visual appearance of many high-resolution observations suggests, especially in active regions.} \citetads{2015SoPh..290.1607S} draw similar conclusion from the analysis of Milne-Eddington inversions of a \ion{He}{i}~$\lambda 10830$ dataset. 
 
\begin{figure*}
\centering
\includegraphics[width=0.85\textwidth, trim = 0cm 0 4.4cm 0, clip]{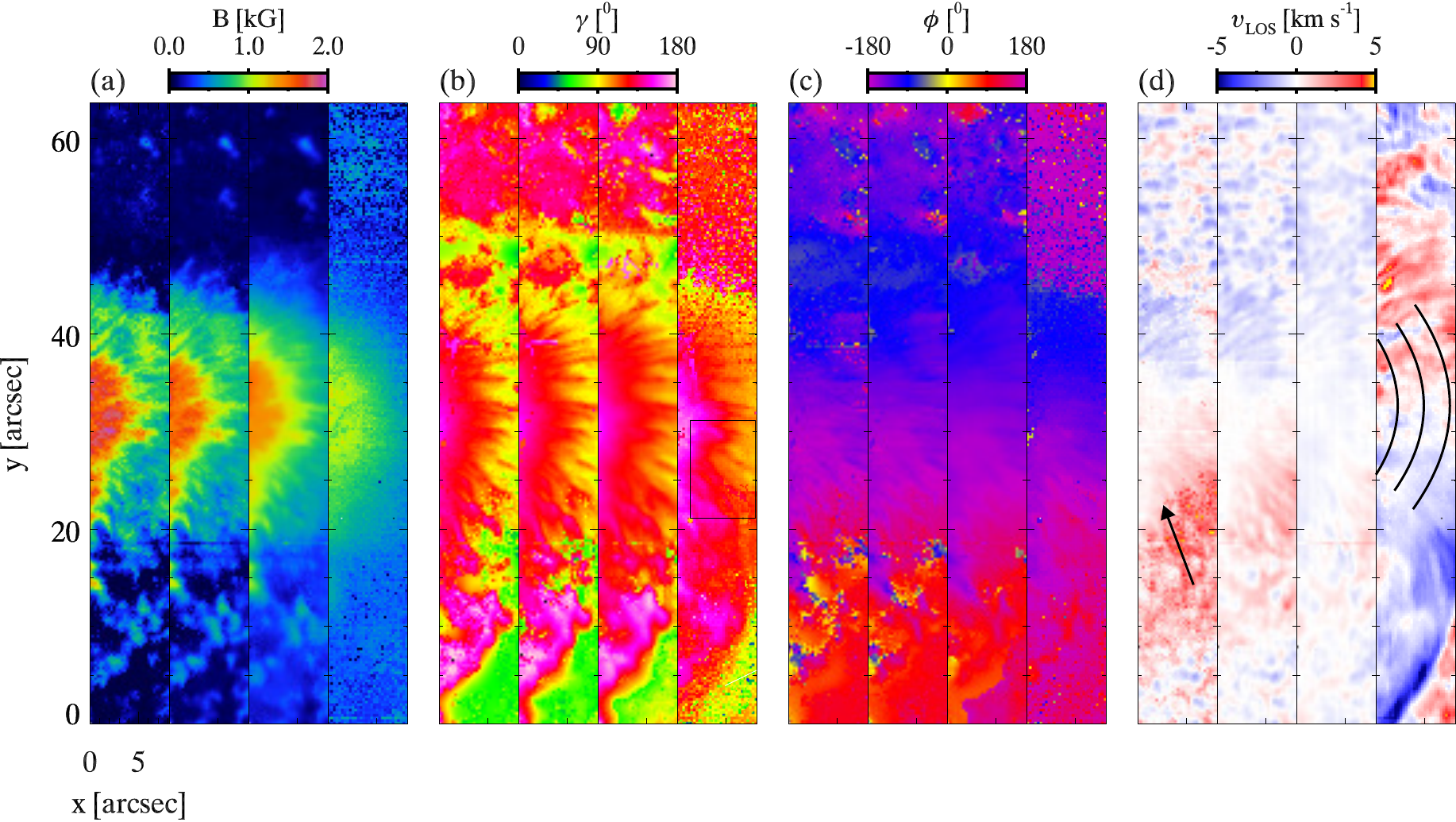}
\caption{Reconstruction of the magnetic field vector from GREGOR/GRIS observations in the \ion{Si}{I}~$\lambda10827$ and the \ion{He}{I}~$\lambda10830$ lines. From left to right there are three groups of panels showing the magnetic field strength, inclination and azimuth. Each group of panels contains the reconstructed quantity as a function of decreasing optical depth from left to right. The first three panels are reconstructed from LTE inversions in the \ion{Si}{I} line. The fourth panel corresponds fo a Milne-Eddington inversion of the \ion{He}{I} line. Reproduced from \citetads{2016arXiv160801988J}, Fig.~3.}\label{fig:incli}
\end{figure*}

 {Very recently, results from the inversion of high-resolution GREGOR/GRIS observations in the \ion{Si}{I}~$\lambda10827$ and the \ion{He}{I}~$\lambda10830$ lines have revealed that the magnetic field strength is very smooth in the upper chromosphere, but the field inclination contains fine structure that is correlated with the inclination of the photospheric magnetic field (see Figure~\ref{fig:incli}, reproduced from \citeads{2016arXiv160801988J})}. 
 
\subsection{An outline for the future}
The future of radiative diagnostics continues to build upon
 forward modelling and inversions. Nowadays, RH (\citeads{2001ApJ...557..389U}), MULTI3D
 (\citeads{2009ASPC..415...87L}) and NICOLE (\citeads{2015A&A...577A...7S}) (among others) are well developed and available to the community. Hanle calculations may be possible with PORTA
 (\citeads{2013A&A...557A.143S}, closed-source) and with TRAVIATA
 (\citeads{2015ApJ...801...16C}, under development). These codes shall be used to develop new
 techniques to accelerate the computation of the NLTE problem (e.g.,
 \citeads{1997A&A...324..161F}; \citeads{2012A&A...543A.109L})
 to implement missing microphysics in the radiative transfer
 calculations (e.g., \citeads{2012JPhCS.388d2018O}), and to study line
 formation with theoretical/numerical model atmospheres.

The methods presented by \citetads{2012A&A...548A...5V} and just
recently by \citetads{2015A&A...577A.140A} {pose the latest development in photospheric inversions. }
These methods couple the parameters of the model
atmosphere spatially and they properly account for the smearing of the telescope
PSF. The application of these methods in the chromosphere will
be a major step forward in the near future, although we have barely
started to exploit the potential of simpler 1.5D inversions in non-LTE. Another point to improve upon is related to the nature of the spatial coupling, which has so far been instrumentally motivated or mathematically imposed, but the usability of the MHD equations remains, to our knowledge, to be proven. 

There is also great potential in the exploration of interesting
chromospheric lines like the \ion{Ca}{ii}~H \& K, or the
\ion{Mg}{ii}~h \& k, although PRD effects must be included to study
these lines. To assess the potential of inversions in H$\alpha$, the
radiative transfer problem must be solved in 3D which largely complicates the
inversion and the description of the model atmosphere.